\documentclass[runningheads]{llncs}
\usepackage{cite}
\usepackage{amsmath,amssymb,amsfonts}
\usepackage{algorithm}
\usepackage{algorithmic}
\usepackage{caption}
\usepackage{graphicx}
\usepackage{textcomp}
\usepackage{multirow}
\usepackage{booktabs}
\usepackage{array}

\def\BibTeX{{\rm B\kern-.05em{\sc i\kern-.025em b}\kern-.08em
    T\kern-.1667em\lower.7ex\hbox{E}\kern-.125emX}}

\begin{document}

\title{A unitary weights based one-iteration quantum perceptron algorithm for non-ideal training sets}
\titlerunning{}
\author{{Wenjie Liu}\inst{1,2} \and PeiPei Gao\inst{2} \and Yuxiang Wang\inst{1,2} \and Wenbin Yu\inst{1,2} \and Maojun Zhang\inst{3,4}}

\institute{Jiangsu Engineering Center of Network Monitoring, Nanjing University of Information Science and Technology, Nanjing 210044, P.R.China\inst{1}\\
School of Computer and Software, Nanjing University of Information Science and Technology, Nanjing 210044, P.R.China\inst{2}\\
School of Mathematics and Computing Science, Guilin University of Electronic Technology, Guilin, China\inst{3}\\
College of Computer Science and Information Technology, Guangxi Normal University, Guilin, China\inst{4}\\
Corresponding author: Maojun Zhang \email{zhang1977108@sina.com}.}

\maketitle

\begin{abstract}
In order to solve the problem of non-ideal training sets (i.e., the less-complete or over-complete sets) and implement one-iteration learning, a novel efficient quantum perceptron algorithm based on unitary weights is proposed, where the singular value decomposition of the total weight matrix from the training set is calculated to make the weight matrix to be unitary. The example validation of quantum gates $\left\{ {H, S, T, \textit{CNOT, Toffoli, Fredkin}} \right\}$ shows that our algorithm can accurately implement arbitrary quantum gates within one iteration. The performance comparison between our algorithm and other quantum perceptron algorithms demonstrates the advantages of our algorithm in terms of applicability, accuracy, and availability. For further validating the applicability of our algorithm, a quantum composite gate which consists of several basic quantum gates is also illustrated.

\end{abstract}


\section{Introduction}
\label{sec:introduction}
With the size of electronic devices becoming smaller and smaller, quantum effects are beginning to interfere in the function of them. The theoretical possibility of quantum computation was firstly explored by Benioff~\cite{Benioff80} and Feynman~\cite{Feynman82}, and the formalization of the quantum computation model was proposed by Deutsch~\cite{Deutsch85}. The main advantage of quantum computation is that the use of superposition and entanglement, which allied with the linearity of the operators, allows for a powerful form of parallelism to devise algorithms more efficient over the known classical ones. There are two well-known representative algorithms, i.e. Shor's factoring algorithm~\cite{Shor99} and Grover's searching algorithm~\cite{Grover97}. The first one is an algorithm for integer factorization formulated, which is NP-hard for classical computation; while Grover searching algorithm achieves a quadratic speedup on the unstructured search spaces over the classical counterparts. After that, the combination of quantum mechanics with other fields (such as information processing, encrypted transmission, machine learning, etc.) has also achieved some important results, such as quantum key agreement (QKA)~\cite{Huang17,Liu18}, quantum secure direct communication (QSDC)~\cite{Liu09,Zhong18}, quantum teleportation and remote state preparation (QT\&RSP)~\cite{Tan18,Liu15,Qu17}, quantum steganography (QS)~\cite{Wu18,Qu18,Qu1802}, delegating quantum computation (DQC)~\cite{Liu17,Liu1802}, and quantum machine learning ~\cite{Lloyd13,Liu1803}.

Artificial neural network (ANN)~\cite{McCulloch43}, also called neural networks (NN), is a computational model used widely in computer science and other research fields. It is based on a large collection of simple neural units (artificial neurons), which is loosely analogous to the observed behavior of a brain's axons. The perceptron~\cite{Rosenblatt58} is the simplest type of neural network classifiers, which only consists of a single artificial neuron.

The debate on quantum approaches to NN (also called quantum neural network, QNN) emerged in the wake of a booming research field of quantum computation two decades ago. In 1995, Kak~\cite{Kak95} put forward some preliminary ideas to find a junction between neural networks and quantum mechanics. Since then, a number of proposals have been proposed~\cite{Lagaris97,Purushothaman97,Andrecut02,Gupta01}. However, in addition to the quantum neural (also called quron), the nonlinear dissipative dynamics of neural computation is fundamentally different from the linear single dynamics of quantum computation~\cite{Nielsen00}. Therefore, how to combine the fields of quantum computation and neural networks is a meaningful task for QNN.

The QNN models can be implemented into four different approaches~\cite{Schuld14}: 1) interpreting the step-function as measurement in order to combine the non-linearity of neural networks and linearity of quantum computation~\cite{Kak95}; 2) using quantum circuits for modeling neural networks~\cite{Panella11}; 3) describing the basic component of ANNs, the perceptron, with a quantum formalism~\cite{Altaisky01}; 4) modeling ANNs with interacting quantum dots consisting of four atoms sharing two electrons~\cite{Behrman99}. In this paper, we focus on the third approach of creating a quantum equivalent of classical perceptron. Thus, it can be seen as a simple unit of QNN that harvests the advantages of quantum information processing.

In 2001, Altaisky~\cite{Altaisky01} firstly proposed a quantum perceptron model to overcome the limitations of the classical perceptron. Although he used the quantum form to represent the classic perceptron model, this proposal is difficult to be extended to a full neural network model. In 2007, Zhou and Ding~\cite{Zhou07} proposed a quantum M-P neural network that could compute the \textit{XOR}-function with one neuron, but it does not follow a unitary evolution and the neuron can be efficiently simulated by a classical single layer neural network. After that, an autonomous quantum perceptron algorithm was proposed by Sagheer and Zidan~\cite{Sagheer13} in 2013, and this algorithm has an iterative rule that can learn to solve a problem in a smaller number of iterations than Altaisky's algorithm. However, its learning weights are non-unitary. Different from the previous models, Siomau~\cite{Siomau14} introduced an autonomous quantum perceptron based on a set of positive operator-valued measures (POVMs). However, the perceptron could not implement basic quantum gates, such as \textit{NOT} and \textit{Hadamard} gates, which are essential for quantum computation.

In order to solve the non-unitary problem, Seow et al.~\cite{Seow15} proposed a new efficient quantum perceptron algorithm in 2015. It expels the iterative learning rule by analytically calculating parameters while maintaining unitary requirements. But the perceptron cannot realize arbitrary quantum computation, and the training set is ideal (i.e, the complete training set). Different from the aforementioned quantum perceptron models~\cite{Altaisky01,Zhou07,Sagheer13,Siomau14,Seow15}, a novel unitary weights based quantum perceptron algorithm is proposed in this study, which not only considers the case of non-ideal training sets, but also implements quantum gates $\left\{ {H, S, T, {\kern 1pt} {\kern 1pt}\textit{CNOT, Toffoli, Fredkin}} \right\}$ (i.e., any quantum computation) correctly within one iteration. For further validating the proposed algorithm, a quantum composite gate which consists of several basic quantum gates is also illustrated.

The remainder of this paper is divided into 5 sections. In Sect.~\ref{sec:2}, some preliminaries including quantum computation, classical perceptron model and singular value decomposition, are briefly introduced. In Sect.~\ref{sec:3}, a unitary weights based quantum perceptron algorithm for non-ideal training sets is proposed, and it can realize general quantum computation within one iteration. In Sect.~\ref{sec:4}, the example validations of some quantum gates $\left\{ {H, S, T, \textit{CNOT}}\right\}$ and $\left\{ \textit{{Toffoli, Fredkin}}\right\}$ are given in detail, and the performance evaluation of our algorithm is conducted in Sect.~\ref{sec:5}. Finally, some remarks and future work are concluded in Sect.~\ref{sec:6}.

\section{Preliminaries}
\label{sec:2}
\subsection{Quantum computation}
As we know, the classic bit is the smallest unit in classic information, and its value is either 0 or 1. In quantum computation, $qubit$ (quantum bit) is quantum analogue of the classic bit, but has two possible values $\left| 0 \right\rangle $ and $\left| 1 \right\rangle $ with a certain probability,
\begin{equation}
  \left| \varphi  \right\rangle  = \alpha \left| 0 \right\rangle  + \beta \left| 1 \right\rangle,
\label{1}
\end{equation}
where $\left| {{\alpha ^2}} \right|{\rm{ + }}{\left| \beta  \right|^2}{\rm{ = }}1$, $\alpha ,\beta$ are complex numbers. Since the vectors $\left| 0 \right\rangle $ and $\left| 1 \right\rangle $ can be represented as follows,
\begin{equation}
  \left| 0 \right\rangle  = \left( \begin{array}{l}
1\\
0
\end{array} \right),  {\kern 1pt} {\kern 1pt} {\kern 1pt} {\kern 1pt} {\kern 1pt} {\kern 1pt} {\kern 1pt} {\kern 1pt} {\kern 1pt} {\kern 1pt} {\kern 1pt} {\kern 1pt} {\kern 1pt} {\kern 1pt} {\kern 1pt} \left| 1 \right\rangle  = \left( \begin{array}{l}
0\\
1
\end{array} \right).
\label{2}
\end{equation}
The qubit $\left| \varphi  \right\rangle $ can be expressed in a vector form $\left| \varphi  \right\rangle  = \left( \begin{smallmatrix}
\alpha \\
\beta
\end{smallmatrix} \right)$.

Quantum operators over a qubit are all represented by $2\times2$ unitary matrices. An $n \times n$ matrix $U$ is unitary if $U{U^\dag } = {U^\dag }U = I$, where ${U^\dag }$ is the conjugate transpose of $U$. For instance, $H$ (\textit{Hadamard)}, $S$ (\textit{phase}), and $T$ ($\pi /8$) operators are important quantum operators over one qubit, and they can be described as $2\times2$ unitary matrices as below,
\begin{equation}
H  =  \frac{1}{{\sqrt 2 }}\left( {\begin{smallmatrix}
1&1\\
1&{ - 1}
\end{smallmatrix}} \right),{\kern 4pt}  S = \left( {\begin{smallmatrix}
1 & 0  \\
0 & i
\end{smallmatrix}} \right),{\kern 4pt} T = \left( {\begin{smallmatrix}
1 & 0  \\
0 & {{e^{i\pi /4}}}
\end{smallmatrix}} \right).
\label{3}
\end{equation}

The \textit{CNOT} gate operates on a quantum register consisting of 2 qubits, known as the control qubit and the target qubit. If the control qubit is set to $0$, then the target qubit is left alone. If the control qubit is set to $1$, then the target qubit is flipped, and \textit{CNOT} operator is described in Eq.~\ref{4},
\begin{equation}
  \textit{CNOT}  = \left( {\begin{array}{*{20}{c}}
1&0&0&0\\
0&1&0&0\\
0&0&0&1\\
0&0&1&0
\end{array}} \right)
\label{4}
\end{equation}

\begin{figure}
\includegraphics[height=3cm ,width=8cm]{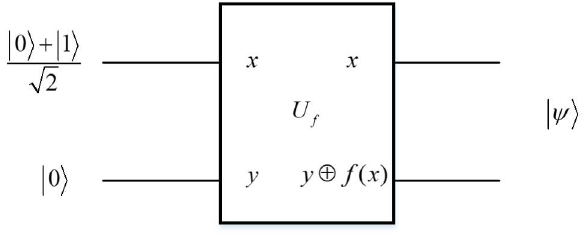}
\caption{Quantum circuit for evaluating $f(0)$ and $f(1)$ simultaneously.}\label{fig:1}
\end{figure}

Quantum parallelism~\cite{Silva16} is a fundamental physical property used in many quantum algorithms. In theory, quantum parallelism, which allows quantum computers to evaluate a function $f(x)$ for many different values of $x$ simultaneously, makes the computational power of quantum computers grow exponentially with the increase in the number of quantum bits. It works as follows, suppose $f(x):{\{ 0,1\}} \to {\{ 0,1\}}$ is a function with one-bit domain and one-bit range, we define the transformation ${U_f}$: $\left| {x,y} \right\rangle  \to \left| {x,y \oplus f(x)} \right\rangle $ (shown in Fig.~\ref{fig:1}), where $\oplus$ indicates addition modulo 2, the first register is called the `data' register, and the second register is the `target' register. In some sense, the quantum parallelism enables the evaluation of all the function $f(x)$ at the same time, even though we only evaluate $f(x)$ once.

\subsection{Classical perceptron}
\begin{figure}
\centering
\includegraphics[height=4cm ,width=5cm]{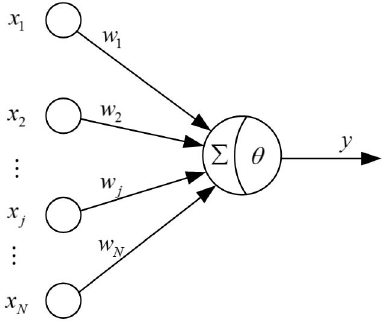}
\caption{The classical perceptron model.}
\label{fig:2}
\end{figure}

The perceptron~\cite{Rosenblatt58} is the simplest type of neural network classifiers, which is an algorithm for learning a binary classifier.
It consists of $N$ input nodes called neurons with values ${x_j} = \{ 1, - 1\} ,{\kern 1pt} {\kern 1pt} {\kern 1pt} j = {\rm{1}},...,{\rm{N}}$, that feed signals into a single output neuron $y$ (shown in Fig.~\ref{fig:2}). Each input neuron is connected to the output neuron with a certain strength denoted by a weight parameter ${w_j} \in [ - 1,{\kern 1pt} {\kern 1pt} {\kern 1pt} 1]$. The input-output relation is governed by the activation function:
\begin{equation}
\hat y = {\rm{f}}\left( {\sum\limits_{j = 1}^N {{w_j}{x_j} - \theta } } \right),
\label{5}
\end{equation}
where the bias $\theta  \in \left[ {{\rm{ - }}1,1} \right]$, and $f(\cdot)$ is the activation function. The perceptron learning algorithm works is shown in Algorithm ~\ref{alg:1}.
\begin{algorithm}
\footnotesize
\caption{The classical perceptron learning algorithm }
\label{alg:1}
\begin{algorithmic}[1]
    \REQUIRE Given a training set in the form $\{ ({x_1},{y_1}),({x_2},{y_2}), \cdots ,({x_N},{y_N})\} $, where ${x_j}$ is an input and ${y_j}$ is the desired output.
    \STATE The weights ${w_j}$ and bias $\theta$ are initialized to small random numbers.
    \STATE The output $\hat y$ generated according to Eq.~\ref{5}.
    \STATE The weights are updated according to the rule ${w_k}(t + {\rm{ 1}}){\rm{ }} = {w_k}(t){\rm{ }} + \eta ( y  - \hat y)x$, where the $\eta$ is learning rate.
    \STATE If $\hat y   = y$, the perceptron does not change. Otherwise go to Step 3 to continue  training and learning.
\end{algorithmic}
\end{algorithm}

\subsection{Singular value decomposition}
In linear algebra, the singular value decomposition(SVD) is a factorization of a real or complex matrix. It is the generalization of the eigendecomposition of a positive semidefinite normal matrix to any matrix via an extension of the polar decomposition. Any rectangular $m \times n$ real or complex matrix $M$ can be decomposed into three matrices and they satisfy:
\begin{equation}
  M = U\Sigma {V^*},
\label{6}
\end{equation}
where $U$ is an $m\times m$ real or complex unitary matrix, $\Sigma$ is an $m\times n$ rectangular diagonal matrix with non-negative real numbers on the diagonal, and $V$ is an $n\times n$ real or complex unitary matrix. The diagonal entries ${\sigma _i}$ of $\Sigma$ are known as the singular values of $M$. The columns of $U$ and the columns of $V$ are called the left-singular vectors and right-singular vectors of $M$, respectively. Applications that employ the SVD include computing the pseudoinverse, least squares fitting of data, multivariable control, matrix approximation, and determining the rank, range and null space of a matrix.

Before we introduce our quantum perceptron algorithm in next section, the key notations and descriptions is firstly listed in Table~\ref{tab:1}.

\begin{table}[!t]
\centering
\footnotesize
\caption{Key notations and descriptions involved in the article.}
\label{tab:1}
\scalebox{1}[1]{
\tabcolsep 5pt 
\begin{tabular}{m{2cm}<{\centering}|m{6cm}<{\centering}}
\toprule
   Notation &Description    \\\hline
  $\left| \varphi  \right\rangle$ & Vector, also known as a $ket$. \\\hline
  $\left\langle \varphi  \right|$ & Vector dual to $\left| \varphi  \right\rangle$, also known as a $bra$. \\\hline
  $\left|\varphi\right\rangle \otimes \left|\psi\right\rangle$ & The tensor product of $\left|\varphi\right\rangle$ and $\left|\psi\right\rangle$, which can be abbreviated as $\left|\varphi\right\rangle \left|\psi\right\rangle$ \\\hline
  ${A^*}$ & Complex conjugate of the $A$ matrix. \\\hline
  ${A^\dag }$ &  Conjugate-transpose of the $A$ matrix, ${A^\dag } = {({A^T})^*}$ \\\hline

  ${\hat w_j}$ & The $j$-th weight defined in Eq.~\ref{8} \\\hline
  ${\hat w}$ & The total weight defined in Eq.~\ref{9}\\\hline
  $\left|y_{output} \right\rangle$ & The output of our perceptron algorithm, defined in Eq.~\ref{11}\\\hline
  $H$ & \textit{Hadmard} gate  \\\hline
  $S$ & \textit{Phase} gate  \\\hline
  $T$ & $\pi /8$ gate  \\\hline
  \textit{CNOT} & Controlled-\emph{NOT} gate  \\\hline
  $\textit{Toffoli}$ & The three-qubit controlled-controlled-\emph{NOT} gate, also named \emph{CCNOT} gate, defined in Eq.~\ref{12} and Fig.~\ref{fig:3} \\\hline
  $\textit{Fredkin}$ & The three-qubit controlled-\emph{SWAP} gate, also named \emph{CSWAP} gate, defined in Eq.~\ref{13} and Fig.~\ref{fig:4}\\
\bottomrule
\end{tabular}
}
\end{table}

\section{The one-iteration quantum perceptron algorithm based on unitary weights}
\label{sec:3}
In order to solve the problem of non-ideal training sets (i.e., the less-complete or over-complete sets) and implement one-iteration learning, we propose a quantum perceptron algorithm based on unitary weights. The algorithm mainly includes: 1) find the conjugate transpose of input, 2) get the total weight, 3) use SVD to decompose weight matrix, 4) get the form of the quantum perceptron. Algorithm 2 shows the detailed procedure of our quantum perceptron algorithm.
\begin{algorithm}
\footnotesize
\caption{Quantum perceptron learning algorithm based on unitary weights}
\label{alg:2}
\begin{algorithmic}[1]
    \REQUIRE Given a training set in the form $\left\{ {\left( {\left| {{x_1}} \right\rangle ,\left| {{y_1}} \right\rangle } \right),\left( {\left| {{x_2}} \right\rangle ,\left| {{y_2}} \right\rangle } \right), \ldots ,\left( {\left| {{x_N}} \right\rangle ,\left| {{y_N}} \right\rangle } \right)} \right\}$, where $\left| {{x_j}} \right\rangle $ is an input and $\left| {{y_j}} \right\rangle $ is the desired output.
    \STATE Find the conjugate transpose of input ${\left| {{x_j}} \right\rangle ^ \dag  } = \left\langle {{x_j}} \right|$.
    \STATE Calculate $\mathop {{\hat w_j}}  = \left| {{y_j}} \right\rangle  \otimes \left\langle {{x_j}} \right|$.
    \STATE Get the total weight $\mathop {\hat w} = \sum\limits_{j = 1}^N {\hat w_j}$.
    \STATE Decompose $\mathop {\hat w}$ into $\mathop {\hat F}\Sigma {\mathop {\hat w_{new}}  }$.
    \STATE Pull out the diagonal matrix of singular values and replace it with a unitary matrix ${\Sigma _{new}}$ with ones in the diagonals and zeroes elsewhere.
    \STATE Get the result of quantum perceptron algorithm $\left| {{y_{output}}} \right\rangle  = \mathop {\hat F} {\Sigma _{new}}\mathop {{\hat w_{new}}} \left| {{x_j}} \right\rangle $.
\end{algorithmic}
\end{algorithm}

\subsection{Find the conjugate transpose of input}
At the beginning of the algorithm, given a training set in the form $\left\{ {\left( {\left| {{x_1}} \right\rangle ,\left| {{y_1}} \right\rangle } \right),\ldots ,\left( {\left| {{x_N}} \right\rangle ,\left| {{y_N}} \right\rangle } \right)} \right\}$, where $\left| {{x_j}} \right\rangle $ is an input and $\left| {{y_j}} \right\rangle $ is the desired output. We can then find the conjugate transpose of input $\left| {{x_j}} \right\rangle$,
\begin{equation}
        {\left| {{x_j}} \right\rangle ^\dag } = \left\langle {{x_j}} \right|.
        \label{7}
        \end{equation}

\subsection{Get the total weight}
After finding the conjugate transpose of input, the main work of this phase is to get the total weight. We take the tensor product of $\left\langle {{x_j}} \right|$ and $\left| {{y_j}} \right\rangle $ to create its corresponding weight ${\hat w_j}$,
    \begin{equation}
        {\mathop {{\hat w_j}}} = \left| {{y_j}} \right\rangle  \otimes \left\langle {{x_j}} \right|,
        \label{8}
        \end{equation}
where $\otimes $ is the tensor product. After calculating the weights, we sum up all of the individual weights, and get the total weight:
    \begin{equation}
        \mathop {\hat w}   = \sum\limits_{j = 1}^N {\hat w_j}.
        \label{9}
        \end{equation}

\subsection{Use SVD to decompose weight matrix}
In order to preserve the quantum properties of being unitary, we decompose non-unitary matrix $\hat w$ into three unitary matrices using SVD. Then we can get two unitary matrices ($\hat F$ and $\hat w_{new}$) and a diagonal matrix $\Sigma$ of singular values as below,
\begin{equation}
    \mathop {\hat w} =\mathop {\hat F} \Sigma {\mathop {\hat w_{new}} }.
    \label{10}
\end{equation}
\subsection{Get the form of quantum perceptron}
Without loss of generality, we remove the diagonal matrix and replace it with a unitary matrix ${\Sigma _{new}}$ of ones in the diagonals and zeros everywhere else, then get the form of quantum perceptron as follows:
\begin{equation}
        \left| {{y_{output}}} \right\rangle  = \mathop {\hat F}  {\Sigma _{new}}\mathop {{\hat w_{new}}} \left| {{x_j}} \right\rangle.
        \label{11}
\end{equation}

\section{Example validation of quantum perceptron algorithm}
\label{sec:4}
As we know, a quantum computer is built from a quantum circuit containing wires and elementary quantum gates to carry around and manipulate the quantum information~\cite{Nielsen00}. Therefore, quantum computation is closely related to the implementation of quantum gates. In order to validate that our proposed algorithm can implement general quantum gates (i.e., quantum computation), the universal quantum gate set $\left\{ {H, S, T,{\kern 1pt} {\kern 1pt} \textit{CNOT}} \right\}$, and some complex quantum gates like \textit{Toffoli} and \textit{Fredkin} gates are illustrated as follows.

\subsection{Universal quantum gate set}
The standard set of universal gates consists of the \textit{Hadamard}, \textit{phase}, $\pi /8$ and \textit{CNOT} gates~\cite{Nielsen00}, and our perceptron algorithm can realize the function of these universal quantum gates within one iteration, the following is the algorithm validation of the universal gates.
\subsubsection{\textit{Hadamard} gate}
\textbf{Example 1} (over-complete) Suppose
\begin{equation}
\nonumber
\begin{array}{l}
\left| {{x_1}} \right\rangle  = \left| 0 \right\rangle ,{\kern 1pt} {\kern 1pt} {\kern 1pt} {\kern 1pt} {\kern 1pt} {\kern 1pt} {\kern 1pt} \left| {{y_1}} \right\rangle  = \frac{{\left| 0 \right\rangle  + \left| 1 \right\rangle }}{{\sqrt 2 }}\\
\left| {{x_2}} \right\rangle  = \left| 1 \right\rangle ,{\kern 1pt} {\kern 1pt} {\kern 1pt} {\kern 1pt} {\kern 1pt} {\kern 1pt} {\kern 1pt} \left| {{y_2}} \right\rangle  = \frac{{\left| 0 \right\rangle  - \left| 1 \right\rangle }}{{\sqrt 2 }}\\
\left| {{x_3}} \right\rangle  = \frac{{{\rm{a}}\left| 0 \right\rangle  + b\left| 1 \right\rangle }}{{\sqrt {{a^2} + {b^2}} }},{\kern 1pt} {\kern 1pt} {\kern 1pt} {\kern 1pt} {\kern 1pt} {\kern 1pt} \left| {{y_3}} \right\rangle  = \frac{{{\rm{(a + b)}}\left| 0 \right\rangle  + (a - b)\left| 1 \right\rangle }}{{\sqrt {2({a^2} + {b^2})} }}(a = 1,b = 2)
\end{array}.
\end{equation}
\textit{Perceptron training:} According to Algorithm~\ref{alg:2}, firstly, the conjugate transpose of inputs are calculated respectively by Eq.~\ref{7}, and the weights are solved by Eq.~\ref{8},
\begin{equation}
\nonumber
\begin{array}{l}
\left\langle {{x_1}} \right| = \left( {1,0} \right){\kern 1pt} {\kern 1pt} {\kern 1pt} \;{\kern 1pt} {\mathop {{\hat w_1}} } = \left| {{y_1}} \right\rangle  \otimes \left\langle {{x_1}} \right| = \frac{1}{{\sqrt 2 }}\left( {\begin{array}{*{20}{c}}
1&0\\
1&0
\end{array}} \right)\\
\left\langle {{x_2}} \right| = \left( {0,1} \right){\kern 1pt} {\kern 1pt} \;{\kern 1pt} {\mathop {{\hat w_2}}} =  \left| {{y_2}} \right\rangle  \otimes \left\langle {{x_2}} \right| = \frac{1}{{\sqrt 2 }}\left( {\begin{array}{*{20}{c}}
0&1\\
0&{ - 1}
\end{array}} \right)\\
\left\langle {{x_3}} \right| = \frac{1}{{\sqrt 5 }}\left( {1,1} \right){\kern 1pt} {\kern 1pt} \;{\kern 1pt} {\kern 1pt} {\mathop {{\hat w_3}} } =\left| {{y_3}} \right\rangle  \otimes \left\langle {{x_3}} \right| = \frac{1}{{\sqrt 2 }}\left( {\begin{array}{*{20}{c}}
{0.6}&{1.2}\\
{ - 0.2}&{ - 0.4}
\end{array}} \right)
\end{array}
\end{equation}
In Step 3, the total weight $\hat w$ can be calculated as follows,
\begin{equation}
\nonumber
\begin{array}{l}
\hat w = \sum\limits_{j = 1}^3 {{{\hat w}_j}} \\
{\kern 1pt} {\kern 1pt} {\kern 1pt} {\kern 1pt} {\kern 1pt} {\kern 1pt}  = \frac{1}{{\sqrt 2 }}\left( {\begin{array}{*{20}{c}}
1&0\\
1&0
\end{array}} \right) + \frac{1}{{\sqrt 2 }}\left( {\begin{array}{*{20}{c}}
0&1\\
0&{ - 1}
\end{array}} \right) + \frac{1}{{\sqrt 2 }}\left( {\begin{array}{*{20}{c}}
{0.6}&{1.2}\\
{ - 0.2}&{ - 0.4}
\end{array}} \right)\\
{\kern 1pt} {\kern 1pt} {\kern 1pt} {\kern 1pt} {\kern 1pt} {\kern 1pt}  = \frac{1}{{\sqrt 2 }}\left( {\begin{array}{*{20}{c}}
{1.6}&{2.2}\\
{0.8}&{ - 1.4}
\end{array}} \right)
\end{array}.
\end{equation}
And in Step 4, SVD is performed on the weight matrix $\hat w$:
\begin{equation}
\nonumber
\begin{array}{l}
\hat F = \frac{1}{{\sqrt 5 }}\left( {\begin{array}{*{20}{c}}
{{{ - 3} \mathord{\left/
 {\vphantom {{ - 3} {\sqrt 2 }}} \right.
 \kern-\nulldelimiterspace} {\sqrt 2 }}}&{{1 \mathord{\left/
 {\vphantom {1 {\sqrt 2 }}} \right.
 \kern-\nulldelimiterspace} {\sqrt 2 }}}\\
{{1 \mathord{\left/
 {\vphantom {1 {\sqrt 2 }}} \right.
 \kern-\nulldelimiterspace} {\sqrt 2 }}}&{{3 \mathord{\left/
 {\vphantom {3 {\sqrt 2 }}} \right.
 \kern-\nulldelimiterspace} {\sqrt 2 }}}
\end{array}} \right),\Sigma  = \left( {\begin{array}{*{20}{c}}
2&0\\
0&1
\end{array}} \right),\\
{{\hat w}_{new}} = \frac{1}{{\sqrt 5 }}\left( {\begin{array}{*{20}{c}}
{ - 1}&{ - 2}\\
2&{ - 1}
\end{array}} \right).
\end{array}
\end{equation}
After the process of amending, we can obtain:
\begin{equation}
\nonumber
\begin{array}{l}
\hat F = \frac{1}{{\sqrt 5 }}\left( {\begin{array}{*{20}{c}}
{{{ - 3} \mathord{\left/
 {\vphantom {{ - 3} {\sqrt 2 }}} \right.
 \kern-\nulldelimiterspace} {\sqrt 2 }}}&{{1 \mathord{\left/
 {\vphantom {1 {\sqrt 2 }}} \right.
 \kern-\nulldelimiterspace} {\sqrt 2 }}}\\
{{1 \mathord{\left/
 {\vphantom {1 {\sqrt 2 }}} \right.
 \kern-\nulldelimiterspace} {\sqrt 2 }}}&{{3 \mathord{\left/
 {\vphantom {3 {\sqrt 2 }}} \right.
 \kern-\nulldelimiterspace} {\sqrt 2 }}}
\end{array}} \right),{\Sigma _{new}} = \left( {\begin{array}{*{20}{c}}
1&0\\
0&1
\end{array}} \right),\\
{{\hat w}_{new}} = \frac{1}{{\sqrt 5 }}\left( {\begin{array}{*{20}{c}}
{ - 1}&{ - 2}\\
2&{ - 1}
\end{array}} \right).
\end{array}
\end{equation}
\textit{Validation of correctness}: The obtained quantum perceptron is used to validate the various inputs. When the perceptron predicting correctly, the output meets $\left| {{y_{output}}} \right\rangle  = \left| {{y_j}} \right\rangle $.
\begin{equation}
\nonumber
\begin{array}{l}
\left| {{y_{output}}} \right\rangle  = \hat F{\Sigma _{new}}{{\hat w}_{new}}\left| {{x_1}} \right\rangle \\
{\kern 1pt} {\kern 1pt} {\kern 1pt} {\kern 1pt} {\kern 1pt} {\kern 1pt} {\kern 1pt} {\kern 1pt} {\kern 1pt} {\kern 1pt} {\kern 1pt} {\kern 1pt} {\kern 1pt} {\kern 1pt} {\kern 1pt} {\kern 1pt} {\kern 1pt} {\kern 1pt} {\kern 1pt} {\kern 1pt} {\kern 1pt} {\kern 1pt} {\kern 1pt} {\kern 1pt} {\kern 1pt} {\kern 1pt} {\kern 1pt} {\kern 1pt} {\kern 1pt} {\kern 1pt} {\kern 1pt} {\kern 1pt} {\kern 1pt} {\kern 1pt} {\kern 1pt}  = \frac{1}{{\sqrt 5 }}\left( {\begin{array}{*{20}{c}}
{{{{\rm{ - }}3} \mathord{\left/
 {\vphantom {{{\rm{ - }}3} {\sqrt 2 }}} \right.
 \kern-\nulldelimiterspace} {\sqrt 2 }}}&{{1 \mathord{\left/
 {\vphantom {1 {\sqrt 2 }}} \right.
 \kern-\nulldelimiterspace} {\sqrt 2 }}}\\
{{1 \mathord{\left/
 {\vphantom {1 {\sqrt 2 }}} \right.
 \kern-\nulldelimiterspace} {\sqrt 2 }}}&{{3 \mathord{\left/
 {\vphantom {3 {\sqrt 2 }}} \right.
 \kern-\nulldelimiterspace} {\sqrt 2 }}}
\end{array}} \right)\left( {\begin{array}{*{20}{c}}
1&0\\
0&1
\end{array}} \right) \times \\
{\kern 1pt} {\kern 1pt} {\kern 1pt} {\kern 1pt} {\kern 1pt} {\kern 1pt} {\kern 1pt} {\kern 1pt} {\kern 1pt} {\kern 1pt} {\kern 1pt} {\kern 1pt} {\kern 1pt} {\kern 1pt} {\kern 1pt} {\kern 1pt} {\kern 1pt} {\kern 1pt} {\kern 1pt} {\kern 1pt} {\kern 1pt} {\kern 1pt} {\kern 1pt} {\kern 1pt} {\kern 1pt} {\kern 1pt} {\kern 1pt} {\kern 1pt} {\kern 1pt} {\kern 1pt} {\kern 1pt} {\kern 1pt} {\kern 1pt} {\kern 1pt} {\kern 1pt} {\kern 1pt} {\kern 1pt} {\kern 1pt} {\kern 1pt} {\kern 1pt} {\kern 1pt} {\kern 1pt} {\kern 1pt} {\kern 1pt} {\kern 1pt} {\kern 1pt} {\kern 1pt} {\kern 1pt} {\kern 1pt} {\kern 1pt} {\kern 1pt} {\kern 1pt} {\kern 1pt} {\kern 1pt} \frac{1}{{\sqrt 5 }}\left( {\begin{array}{*{20}{c}}
{ - 1}&{ - 2}\\
2&{ - 1}
\end{array}} \right)\left( {\begin{array}{*{20}{c}}
1\\
0
\end{array}} \right)\\
{\kern 1pt} {\kern 1pt} {\kern 1pt} {\kern 1pt} {\kern 1pt} {\kern 1pt} {\kern 1pt} {\kern 1pt} {\kern 1pt} {\kern 1pt} {\kern 1pt} {\kern 1pt} {\kern 1pt} {\kern 1pt} {\kern 1pt} {\kern 1pt} {\kern 1pt} {\kern 1pt} {\kern 1pt} {\kern 1pt} {\kern 1pt} {\kern 1pt} {\kern 1pt} {\kern 1pt} {\kern 1pt} {\kern 1pt} {\kern 1pt} {\kern 1pt} {\kern 1pt} {\kern 1pt} {\kern 1pt} {\kern 1pt} {\kern 1pt} {\kern 1pt}  = \frac{{\left| 0 \right\rangle  + \left| 1 \right\rangle }}{{\sqrt 2 }}\\
{\kern 1pt} {\kern 1pt} {\kern 1pt} {\kern 1pt} {\kern 1pt} {\kern 1pt} {\kern 1pt} {\kern 1pt} {\kern 1pt} {\kern 1pt} {\kern 1pt} {\kern 1pt} {\kern 1pt} {\kern 1pt} {\kern 1pt} {\kern 1pt} {\kern 1pt} {\kern 1pt} {\kern 1pt} {\kern 1pt} {\kern 1pt} {\kern 1pt} {\kern 1pt} {\kern 1pt} {\kern 1pt} {\kern 1pt} {\kern 1pt} {\kern 1pt} {\kern 1pt} {\kern 1pt} {\kern 1pt} {\kern 1pt} {\kern 1pt}  = \left| {{y_1}} \right\rangle \\
\left| {{y_{output}}} \right\rangle  = \hat F{\Sigma _{new}}{{\hat w}_{new}}\left| {{x_2}} \right\rangle \\
{\kern 1pt} {\kern 1pt} {\kern 1pt} {\kern 1pt} {\kern 1pt} {\kern 1pt} {\kern 1pt} {\kern 1pt} {\kern 1pt} {\kern 1pt} {\kern 1pt} {\kern 1pt} {\kern 1pt} {\kern 1pt} {\kern 1pt} {\kern 1pt} {\kern 1pt} {\kern 1pt} {\kern 1pt} {\kern 1pt} {\kern 1pt} {\kern 1pt} {\kern 1pt} {\kern 1pt} {\kern 1pt} {\kern 1pt} {\kern 1pt} {\kern 1pt} {\kern 1pt} {\kern 1pt} {\kern 1pt} {\kern 1pt} {\kern 1pt} {\kern 1pt} {\kern 1pt}  = \frac{1}{{\sqrt 5 }}\left( {\begin{array}{*{20}{c}}
{{{{\rm{ - }}3} \mathord{\left/
 {\vphantom {{{\rm{ - }}3} {\sqrt 2 }}} \right.
 \kern-\nulldelimiterspace} {\sqrt 2 }}}&{{1 \mathord{\left/
 {\vphantom {1 {\sqrt 2 }}} \right.
 \kern-\nulldelimiterspace} {\sqrt 2 }}}\\
{{1 \mathord{\left/
 {\vphantom {1 {\sqrt 2 }}} \right.
 \kern-\nulldelimiterspace} {\sqrt 2 }}}&{{3 \mathord{\left/
 {\vphantom {3 {\sqrt 2 }}} \right.
 \kern-\nulldelimiterspace} {\sqrt 2 }}}
\end{array}} \right)\left( {\begin{array}{*{20}{c}}
1&0\\
0&1
\end{array}} \right) \times \\
{\kern 1pt} {\kern 1pt} {\kern 1pt} {\kern 1pt} {\kern 1pt} {\kern 1pt} {\kern 1pt} {\kern 1pt} {\kern 1pt} {\kern 1pt} {\kern 1pt} {\kern 1pt} {\kern 1pt} {\kern 1pt} {\kern 1pt} {\kern 1pt} {\kern 1pt} {\kern 1pt} {\kern 1pt} {\kern 1pt} {\kern 1pt} {\kern 1pt} {\kern 1pt} {\kern 1pt} {\kern 1pt} {\kern 1pt} {\kern 1pt} {\kern 1pt} {\kern 1pt} {\kern 1pt} {\kern 1pt} {\kern 1pt} {\kern 1pt} {\kern 1pt} {\kern 1pt} {\kern 1pt} {\kern 1pt} {\kern 1pt} {\kern 1pt} {\kern 1pt} {\kern 1pt} {\kern 1pt} {\kern 1pt} {\kern 1pt} {\kern 1pt} {\kern 1pt} {\kern 1pt} {\kern 1pt} {\kern 1pt} {\kern 1pt} {\kern 1pt} {\kern 1pt} {\kern 1pt} {\kern 1pt} \frac{1}{{\sqrt 5 }}\left( {\begin{array}{*{20}{c}}
{ - 1}&{ - 2}\\
2&{ - 1}
\end{array}} \right)\left( {\begin{array}{*{20}{c}}
0\\
1
\end{array}} \right)\\
{\kern 1pt} {\kern 1pt} {\kern 1pt} {\kern 1pt} {\kern 1pt} {\kern 1pt} {\kern 1pt} {\kern 1pt} {\kern 1pt} {\kern 1pt} {\kern 1pt} {\kern 1pt} {\kern 1pt} {\kern 1pt} {\kern 1pt} {\kern 1pt} {\kern 1pt} {\kern 1pt} {\kern 1pt} {\kern 1pt} {\kern 1pt} {\kern 1pt} {\kern 1pt} {\kern 1pt} {\kern 1pt} {\kern 1pt} {\kern 1pt} {\kern 1pt} {\kern 1pt} {\kern 1pt} {\kern 1pt} {\kern 1pt} {\kern 1pt} {\kern 1pt}  = \frac{{\left| 0 \right\rangle  - \left| 1 \right\rangle }}{{\sqrt 2 }}\\
{\kern 1pt} {\kern 1pt} {\kern 1pt} {\kern 1pt} {\kern 1pt} {\kern 1pt} {\kern 1pt} {\kern 1pt} {\kern 1pt} {\kern 1pt} {\kern 1pt} {\kern 1pt} {\kern 1pt} {\kern 1pt} {\kern 1pt} {\kern 1pt} {\kern 1pt} {\kern 1pt} {\kern 1pt} {\kern 1pt} {\kern 1pt} {\kern 1pt} {\kern 1pt} {\kern 1pt} {\kern 1pt} {\kern 1pt} {\kern 1pt} {\kern 1pt} {\kern 1pt} {\kern 1pt} {\kern 1pt} {\kern 1pt} {\kern 1pt} {\kern 1pt}  = \left| {{y_2}} \right\rangle \\
\left| {{y_{output}}} \right\rangle  = \hat F{\Sigma _{new}}{{\hat w}_{new}}\left| {{x_3}} \right\rangle \\
{\kern 1pt} {\kern 1pt} {\kern 1pt} {\kern 1pt} {\kern 1pt} {\kern 1pt} {\kern 1pt} {\kern 1pt} {\kern 1pt} {\kern 1pt} {\kern 1pt} {\kern 1pt} {\kern 1pt} {\kern 1pt} {\kern 1pt} {\kern 1pt} {\kern 1pt} {\kern 1pt} {\kern 1pt} {\kern 1pt} {\kern 1pt} {\kern 1pt} {\kern 1pt} {\kern 1pt} {\kern 1pt} {\kern 1pt} {\kern 1pt} {\kern 1pt} {\kern 1pt} {\kern 1pt} {\kern 1pt} {\kern 1pt} {\kern 1pt} {\kern 1pt} {\kern 1pt}  = \frac{1}{{\sqrt 5 }}\left( {\begin{array}{*{20}{c}}
{{{{\rm{ - }}3} \mathord{\left/
 {\vphantom {{{\rm{ - }}3} {\sqrt 2 }}} \right.
 \kern-\nulldelimiterspace} {\sqrt 2 }}}&{{1 \mathord{\left/
 {\vphantom {1 {\sqrt 2 }}} \right.
 \kern-\nulldelimiterspace} {\sqrt 2 }}}\\
{{1 \mathord{\left/
 {\vphantom {1 {\sqrt 2 }}} \right.
 \kern-\nulldelimiterspace} {\sqrt 2 }}}&{{3 \mathord{\left/
 {\vphantom {3 {\sqrt 2 }}} \right.
 \kern-\nulldelimiterspace} {\sqrt 2 }}}
\end{array}} \right)\left( {\begin{array}{*{20}{c}}
1&0\\
0&1
\end{array}} \right) \times \\
{\kern 1pt} {\kern 1pt} {\kern 1pt} {\kern 1pt} {\kern 1pt} {\kern 1pt} {\kern 1pt} {\kern 1pt} {\kern 1pt} {\kern 1pt} {\kern 1pt} {\kern 1pt} {\kern 1pt} {\kern 1pt} {\kern 1pt} {\kern 1pt} {\kern 1pt} {\kern 1pt} {\kern 1pt} {\kern 1pt} {\kern 1pt} {\kern 1pt} {\kern 1pt} {\kern 1pt} {\kern 1pt} {\kern 1pt} {\kern 1pt} {\kern 1pt} {\kern 1pt} {\kern 1pt} {\kern 1pt} {\kern 1pt} {\kern 1pt} {\kern 1pt} {\kern 1pt} {\kern 1pt} {\kern 1pt} {\kern 1pt} {\kern 1pt} {\kern 1pt} {\kern 1pt} {\kern 1pt} {\kern 1pt} {\kern 1pt} {\kern 1pt} {\kern 1pt} {\kern 1pt} {\kern 1pt} {\kern 1pt} {\kern 1pt} \frac{1}{{\sqrt 5 }}\left( {\begin{array}{*{20}{c}}
{ - 1}&{ - 2}\\
2&{ - 1}
\end{array}} \right)\left( {\begin{array}{*{20}{c}}
1\\
2
\end{array}} \right)\\
{\kern 1pt} {\kern 1pt} {\kern 1pt} {\kern 1pt} {\kern 1pt} {\kern 1pt} {\kern 1pt} {\kern 1pt} {\kern 1pt} {\kern 1pt} {\kern 1pt} {\kern 1pt} {\kern 1pt} {\kern 1pt} {\kern 1pt} {\kern 1pt} {\kern 1pt} {\kern 1pt} {\kern 1pt} {\kern 1pt} {\kern 1pt} {\kern 1pt} {\kern 1pt} {\kern 1pt} {\kern 1pt} {\kern 1pt} {\kern 1pt} {\kern 1pt} {\kern 1pt} {\kern 1pt} {\kern 1pt} {\kern 1pt} {\kern 1pt} {\kern 1pt}  = \frac{{3\left| 0 \right\rangle  - \left| 1 \right\rangle }}{{\sqrt {2 \times 5} }}\\
{\kern 1pt} {\kern 1pt} {\kern 1pt} {\kern 1pt} {\kern 1pt} {\kern 1pt} {\kern 1pt} {\kern 1pt} {\kern 1pt} {\kern 1pt} {\kern 1pt} {\kern 1pt} {\kern 1pt} {\kern 1pt} {\kern 1pt} {\kern 1pt} {\kern 1pt} {\kern 1pt} {\kern 1pt} {\kern 1pt} {\kern 1pt} {\kern 1pt} {\kern 1pt} {\kern 1pt} {\kern 1pt} {\kern 1pt} {\kern 1pt} {\kern 1pt} {\kern 1pt} {\kern 1pt} {\kern 1pt} {\kern 1pt} {\kern 1pt} {\kern 1pt}  = \left| {{y_3}} \right\rangle
\end{array}
\end{equation}
\textbf{Example 2} (less-complete) Suppose
\begin{equation}
\nonumber
\begin{array}{l}
\left| {{x_1}} \right\rangle  = \frac{{{\rm{a}}\left| 0 \right\rangle  + b\left| 1 \right\rangle }}{{\sqrt {{a^2} + {b^2}} }},\\
\left| {{y_1}} \right\rangle  = \frac{{({\rm{a + b}})\left| 0 \right\rangle  + (a - b)\left| 1 \right\rangle }}{{\sqrt {2({a^2} + {b^2})} }}(a =  - 11,b = 7).
\end{array}
\end{equation}
\textit{Perceptron training}: The same way through Algorithm~\ref{alg:2} for training and finally get the quantum perceptron,
\begin{equation}
\nonumber
\begin{array}{*{20}{l}}
{\left| {{y_{output}}} \right\rangle  = \mathop {{\rm{ }}\hat F}  {\Sigma _{new}}\mathop {{\hat w_{new}}}   \left| {{x_j}} \right\rangle }\\
\begin{array}{l}
 {\kern 1pt} {\kern 1pt} {\kern 1pt} {\kern 1pt} {\kern 1pt} {\kern 1pt} {\kern 1pt} {\kern 1pt} {\kern 1pt} {\kern 1pt} {\kern 1pt} {\kern 1pt} {\kern 1pt} {\kern 1pt} {\kern 1pt} {\kern 1pt} {\kern 1pt} {\kern 1pt} {\kern 1pt} {\kern 1pt} {\kern 1pt} {\kern 1pt} {\kern 1pt} {\kern 1pt} {\kern 1pt} {\kern 1pt} {\kern 1pt} {\kern 1pt} {\kern 1pt} {\kern 1pt} {\kern 1pt}  = \frac{1}{{\sqrt {170} }}\left( {\begin{array}{*{20}{c}}
{ - 2\sqrt 2 }&{ - 9\sqrt 2 }\\
{ - 9\sqrt 2 }&{2\sqrt 2 }
\end{array}} \right)\left( {\begin{array}{*{20}{c}}
1&0\\
0&1
\end{array}} \right) \times \\
 {\kern 1pt} {\kern 1pt} {\kern 1pt} {\kern 1pt} {\kern 1pt} {\kern 1pt} {\kern 1pt} {\kern 1pt} {\kern 1pt} {\kern 1pt} {\kern 1pt} {\kern 1pt} {\kern 1pt} {\kern 1pt} {\kern 1pt} {\kern 1pt} {\kern 1pt} {\kern 1pt} {\kern 1pt} {\kern 1pt} {\kern 1pt} {\kern 1pt} {\kern 1pt} {\kern 1pt} {\kern 1pt} {\kern 1pt} {\kern 1pt} {\kern 1pt} {\kern 1pt} {\kern 1pt} {\kern 1pt} {\kern 1pt} {\kern 1pt} {\kern 1pt} {\kern 1pt} {\kern 1pt} {\kern 1pt} {\kern 1pt} {\kern 1pt} {\kern 1pt} {\kern 1pt} {\kern 1pt} {\kern 1pt} {\kern 1pt} \frac{1}{{\sqrt {170} }}\left( {\begin{array}{*{20}{c}}
{ - 11}&7\\
{ - 7}&{ - 11}
\end{array}} \right)\left| {{x_j}} \right\rangle .
\end{array}
\end{array}
\end{equation}
\textit{Validation of correctness}:
\begin{equation}
\nonumber
\begin{array}{*{20}{l}}
{\left| {{y_{output}}} \right\rangle  = \mathop {\hat F}   {\Sigma _{new}}\mathop {{\hat w_{new}}} \left| {{x_1}} \right\rangle }\\
{\begin{array}{*{20}{l}}
{{\kern 1pt} {\kern 1pt} {\kern 1pt} {\kern 1pt} {\kern 1pt} {\kern 1pt} {\kern 1pt} {\kern 1pt} {\kern 1pt} {\kern 1pt} {\kern 1pt} {\kern 1pt} {\kern 1pt} {\kern 1pt} {\kern 1pt} {\kern 1pt} {\kern 1pt} {\kern 1pt} {\kern 1pt} {\kern 1pt} {\kern 1pt} {\kern 1pt} {\kern 1pt} {\kern 1pt} {\kern 1pt} {\kern 1pt} {\kern 1pt} {\kern 1pt} {\kern 1pt} {\kern 1pt}  = \frac{1}{{\sqrt {170} }}\left( {\begin{array}{*{20}{c}}
{ - 2\sqrt 2 }&{ - 9\sqrt 2 }\\
{ - 9\sqrt 2 }&{2\sqrt 2 }
\end{array}} \right)\left( {\begin{array}{*{20}{c}}
1&0\\
0&1
\end{array}} \right) \times }\\
{{\kern 1pt} {\kern 1pt} {\kern 1pt} {\kern 1pt} {\kern 1pt} {\kern 1pt} {\kern 1pt} {\kern 1pt} {\kern 1pt} {\kern 1pt} {\kern 1pt} {\kern 1pt} {\kern 1pt} {\kern 1pt} {\kern 1pt} {\kern 1pt} {\kern 1pt} {\kern 1pt} {\kern 1pt} {\kern 1pt} {\kern 1pt} {\kern 1pt} {\kern 1pt} {\kern 1pt} {\kern 1pt} {\kern 1pt} {\kern 1pt} {\kern 1pt} {\kern 1pt} {\kern 1pt} {\kern 1pt} {\kern 1pt} {\kern 1pt} {\kern 1pt} {\kern 1pt} {\kern 1pt} {\kern 1pt} {\kern 1pt} {\kern 1pt} \frac{1}{{\sqrt {170} }}\left( {\begin{array}{*{20}{c}}
{ - 11}&7\\
{ - 7}&{ - 11}
\end{array}} \right)\frac{1}{{\sqrt {170} }}\left( {\begin{array}{*{20}{l}}
{ - 11}\\
7
\end{array}} \right)}
\end{array}}\\
\begin{array}{l}
{\kern 1pt} {\kern 1pt} {\kern 1pt} {\kern 1pt} {\kern 1pt} {\kern 1pt} {\kern 1pt} {\kern 1pt} {\kern 1pt} {\kern 1pt} {\kern 1pt} {\kern 1pt} {\kern 1pt} {\kern 1pt} {\kern 1pt} {\kern 1pt} {\kern 1pt} {\kern 1pt} {\kern 1pt} {\kern 1pt} {\kern 1pt} {\kern 1pt} {\kern 1pt} {\kern 1pt} {\kern 1pt} {\kern 1pt} {\kern 1pt} {\kern 1pt} {\kern 1pt} {\kern 1pt}  = \frac{{ - 4\left| 0 \right\rangle  - 18\left| 1 \right\rangle }}{{\sqrt {2 \times 170} }}\\
{\kern 1pt} {\kern 1pt} {\kern 1pt} {\kern 1pt} {\kern 1pt} {\kern 1pt} {\kern 1pt} {\kern 1pt} {\kern 1pt} {\kern 1pt} {\kern 1pt} {\kern 1pt} {\kern 1pt} {\kern 1pt} {\kern 1pt} {\kern 1pt} {\kern 1pt} {\kern 1pt} {\kern 1pt} {\kern 1pt} {\kern 1pt} {\kern 1pt} {\kern 1pt} {\kern 1pt} {\kern 1pt} {\kern 1pt} {\kern 1pt} {\kern 1pt} {\kern 1pt} {\kern 1pt}  = \left| {{y_1}} \right\rangle .
\end{array}
\end{array}
\end{equation}

\subsubsection{\textit{Phase} gate }
\textbf{Example 3} (over-complete) Suppose
\begin{equation}
\nonumber
\begin{array}{l}
\left| {{x_1}} \right\rangle  = \left| 0 \right\rangle ,{\kern 1pt} {\kern 1pt} {\kern 1pt} {\kern 1pt} {\kern 1pt} {\kern 1pt} \left| {{y_1}} \right\rangle  = \left| 0 \right\rangle \\
\left| {{x_2}} \right\rangle  = \left| 1 \right\rangle ,{\kern 1pt} {\kern 1pt} {\kern 1pt} {\kern 1pt} {\kern 1pt} {\kern 1pt} {\kern 1pt} \left| {{y_2}} \right\rangle  = i\left| 1 \right\rangle \\
\left| {{x_3}} \right\rangle  = \frac{{{\rm{a}}\left| 0 \right\rangle  + b\left| 1 \right\rangle }}{{\sqrt {{a^2} + {b^2}} }},{\kern 1pt} {\kern 1pt} {\kern 1pt} {\kern 1pt} {\kern 1pt} {\kern 1pt} \left| {{y_3}} \right\rangle  = \frac{{a\left| 0 \right\rangle  + bi\left| 1 \right\rangle }}{{\sqrt {{a^2} + {b^2}} }}(a = 2,b = 2)
\end{array}
\end{equation}
\textit{Perceptron training}: The quantum perceptron is
\begin{equation}
\nonumber
\begin{array}{*{20}{l}}
{\left| {{y_{output}}} \right\rangle  = \mathop {{\rm{ }}\hat F} {\Sigma _{new}}\mathop {{\hat w_{new}}}  \left| {{x_j}} \right\rangle }\\
\begin{array}{l}
{\kern 1pt} {\kern 1pt} {\kern 1pt} {\kern 1pt} {\kern 1pt} {\kern 1pt} {\kern 1pt} {\kern 1pt} {\kern 1pt} {\kern 1pt} {\kern 1pt} {\kern 1pt} {\kern 1pt} {\kern 1pt} {\kern 1pt} {\kern 1pt} {\kern 1pt} {\kern 1pt} {\kern 1pt} {\kern 1pt} {\kern 1pt} {\kern 1pt} {\kern 1pt} {\kern 1pt} {\kern 1pt} {\kern 1pt} {\kern 1pt} {\kern 1pt} {\kern 1pt} {\kern 1pt}  = \frac{1}{{\sqrt 2 }}\left( {\begin{array}{*{20}{c}}
{ - 1}&1\\
{ - i}&{ - i}
\end{array}} \right)\left( {\begin{array}{*{20}{c}}
1&0\\
0&1
\end{array}} \right) \times \\
{\kern 1pt} {\kern 1pt}{\kern 1pt} {\kern 1pt} {\kern 1pt} {\kern 1pt} {\kern 1pt} {\kern 1pt} {\kern 1pt}  {\kern 1pt}{\kern 1pt} {\kern 1pt} {\kern 1pt} {\kern 1pt} {\kern 1pt} {\kern 1pt} {\kern 1pt}  {\kern 1pt} {\kern 1pt} {\kern 1pt} {\kern 1pt} {\kern 1pt} {\kern 1pt} {\kern 1pt} {\kern 1pt} {\kern 1pt} {\kern 1pt} {\kern 1pt} {\kern 1pt} {\kern 1pt} {\kern 1pt} {\kern 1pt} {\kern 1pt} {\kern 1pt} {\kern 1pt} {\kern 1pt} {\kern 1pt} {\kern 1pt} {\kern 1pt} {\kern 1pt} {\kern 1pt} {\kern 1pt} \frac{1}{{\sqrt 2 }}\left( {\begin{array}{*{20}{c}}
{ - 1}&{ - 1}\\
1&{ - 1}
\end{array}} \right)\left| {{x_j}} \right\rangle .
\end{array}
\end{array}.
\end{equation}
\textit{Validation of correctness}:
\begin{equation}
\nonumber
\begin{array}{l}
\begin{array}{*{20}{l}}
{\left| {{y_{output}}} \right\rangle  = \mathop {\hat F} {\Sigma _{new}}\mathop {{\hat w_{new}}} \left| {{x_1}} \right\rangle }\\
\begin{array}{l}
\begin{array}{*{20}{l}}
{{\kern 1pt} {\kern 1pt} {\kern 1pt}  {\kern 1pt} {\kern 1pt}  {\kern 1pt} {\kern 1pt} {\kern 1pt} {\kern 1pt} {\kern 1pt} {\kern 1pt} {\kern 1pt} {\kern 1pt} {\kern 1pt} {\kern 1pt} {\kern 1pt} {\kern 1pt} {\kern 1pt} {\kern 1pt} {\kern 1pt} {\kern 1pt} {\kern 1pt} {\kern 1pt} {\kern 1pt} {\kern 1pt} {\kern 1pt}  = \frac{1}{{\sqrt 2 }}\left( {\begin{array}{*{20}{c}}
{ - 1}&1\\
{ - i}&{ - i}
\end{array}} \right)\left( {\begin{array}{*{20}{c}}
1&0\\
0&1
\end{array}} \right) \times }\\
\begin{array}{l}
{\kern 1pt} {\kern 1pt} {\kern 1pt}  {\kern 1pt} {\kern 1pt} {\kern 1pt} {\kern 1pt} {\kern 1pt} {\kern 1pt} {\kern 1pt} {\kern 1pt} {\kern 1pt} {\kern 1pt} {\kern 1pt} {\kern 1pt} {\kern 1pt} {\kern 1pt} {\kern 1pt} {\kern 1pt} {\kern 1pt} {\kern 1pt} {\kern 1pt} {\kern 1pt} {\kern 1pt} {\kern 1pt} {\kern 1pt} {\kern 1pt} {\kern 1pt} {\kern 1pt} {\kern 1pt} {\kern 1pt} {\kern 1pt} {\kern 1pt} {\kern 1pt} {\kern 1pt} {\kern 1pt} {\kern 1pt} {\kern 1pt} \frac{1}{{\sqrt 2 }}\left( {\begin{array}{*{20}{c}}
{ - 1}&{ - 1}\\
1&{ - 1}
\end{array}} \right)\left( {\begin{array}{*{20}{c}}
1\\
0
\end{array}} \right)\\
{\kern 1pt}  {\kern 1pt} {\kern 1pt} {\kern 1pt} {\kern 1pt} {\kern 1pt} {\kern 1pt} {\kern 1pt} {\kern 1pt} {\kern 1pt} {\kern 1pt} {\kern 1pt} {\kern 1pt} {\kern 1pt} {\kern 1pt} {\kern 1pt} {\kern 1pt} {\kern 1pt} {\kern 1pt} {\kern 1pt} {\kern 1pt} {\kern 1pt}  = \left| 0 \right\rangle \\
{\kern 1pt}{\kern 1pt} {\kern 1pt} {\kern 1pt} {\kern 1pt} {\kern 1pt} {\kern 1pt} {\kern 1pt} {\kern 1pt} {\kern 1pt} {\kern 1pt} {\kern 1pt} {\kern 1pt} {\kern 1pt} {\kern 1pt} {\kern 1pt} {\kern 1pt} {\kern 1pt} {\kern 1pt} {\kern 1pt} {\kern 1pt} {\kern 1pt}  = \left| {{y_1}} \right\rangle
\end{array}
\end{array}\\
\begin{array}{*{20}{l}}
{\left| {{y_{output}}} \right\rangle  = \mathop {\hat F} {\Sigma _{new}}\mathop {{\hat w_{new}}} \left| {{x_2}} \right\rangle }\\
{\begin{array}{*{20}{l}}
{{\kern 1pt}  {\kern 1pt} {\kern 1pt} {\kern 1pt} {\kern 1pt} {\kern 1pt} {\kern 1pt} {\kern 1pt} {\kern 1pt} {\kern 1pt} {\kern 1pt} {\kern 1pt} {\kern 1pt} {\kern 1pt} {\kern 1pt} {\kern 1pt} {\kern 1pt} {\kern 1pt} {\kern 1pt} {\kern 1pt} {\kern 1pt} {\kern 1pt} {\kern 1pt} {\kern 1pt} {\kern 1pt} {\kern 1pt} {\kern 1pt} {\kern 1pt} {\kern 1pt} {\kern 1pt} {\kern 1pt}  = \frac{1}{{\sqrt 2 }}\left( {\begin{array}{*{20}{c}}
{ - 1}&1\\
{ - i}&{ - i}
\end{array}} \right)\left( {\begin{array}{*{20}{c}}
1&0\\
0&1
\end{array}} \right) \times }\\
\begin{array}{l}
{\kern 1pt} {\kern 1pt} {\kern 1pt} {\kern 1pt} {\kern 1pt} {\kern 1pt} {\kern 1pt} {\kern 1pt} {\kern 1pt} {\kern 1pt} {\kern 1pt} {\kern 1pt} {\kern 1pt} {\kern 1pt} {\kern 1pt} {\kern 1pt} {\kern 1pt} {\kern 1pt} {\kern 1pt} {\kern 1pt} {\kern 1pt} {\kern 1pt} {\kern 1pt} {\kern 1pt} {\kern 1pt} {\kern 1pt} {\kern 1pt} {\kern 1pt} {\kern 1pt} {\kern 1pt} {\kern 1pt} {\kern 1pt} {\kern 1pt} {\kern 1pt} {\kern 1pt} {\kern 1pt} {\kern 1pt} {\kern 1pt} {\kern 1pt} \frac{1}{{\sqrt 2 }}\left( {\begin{array}{*{20}{c}}
{ - 1}&{ - 1}\\
1&{ - 1}
\end{array}} \right)\left( {\begin{array}{*{20}{c}}
0\\
1
\end{array}} \right)\\
{\kern 1pt} {\kern 1pt} {\kern 1pt} {\kern 1pt} {\kern 1pt} {\kern 1pt} {\kern 1pt} {\kern 1pt} {\kern 1pt} {\kern 1pt} {\kern 1pt} {\kern 1pt} {\kern 1pt} {\kern 1pt} {\kern 1pt} {\kern 1pt} {\kern 1pt} {\kern 1pt} {\kern 1pt} {\kern 1pt} {\kern 1pt} {\kern 1pt} {\kern 1pt} {\kern 1pt} {\kern 1pt} {\kern 1pt} {\kern 1pt}  = i\left| 1 \right\rangle \\
{\kern 1pt} {\kern 1pt} {\kern 1pt} {\kern 1pt} {\kern 1pt} {\kern 1pt} {\kern 1pt} {\kern 1pt} {\kern 1pt} {\kern 1pt} {\kern 1pt} {\kern 1pt} {\kern 1pt} {\kern 1pt} {\kern 1pt} {\kern 1pt} {\kern 1pt} {\kern 1pt} {\kern 1pt} {\kern 1pt} {\kern 1pt} {\kern 1pt} {\kern 1pt} {\kern 1pt} {\kern 1pt} {\kern 1pt} {\kern 1pt}  = \left| {{y_2}} \right\rangle
\end{array}
\end{array}}
\end{array}
\end{array}
\end{array}\\
\begin{array}{*{20}{l}}
{\left| {{y_{output}}} \right\rangle  = \mathop {\hat F} {\Sigma _{new}}\mathop {{\hat w_{new}}}\left| {{x_3}} \right\rangle }\\
{\begin{array}{*{20}{l}}
{{\kern 1pt}  {\kern 1pt} {\kern 1pt} {\kern 1pt} {\kern 1pt} {\kern 1pt} {\kern 1pt} {\kern 1pt} {\kern 1pt} {\kern 1pt} {\kern 1pt} {\kern 1pt} {\kern 1pt} {\kern 1pt} {\kern 1pt} {\kern 1pt} {\kern 1pt} {\kern 1pt} {\kern 1pt} {\kern 1pt} {\kern 1pt} {\kern 1pt} {\kern 1pt} {\kern 1pt} {\kern 1pt} {\kern 1pt} {\kern 1pt} {\kern 1pt} {\kern 1pt} {\kern 1pt} {\kern 1pt}  = \frac{1}{{\sqrt 2 }}\left( {\begin{array}{*{20}{c}}
{ - 1}&1\\
{ - i}&{ - i}
\end{array}} \right)\left( {\begin{array}{*{20}{c}}
1&0\\
0&1
\end{array}} \right) \times }\\
\begin{array}{l}
{\kern 1pt} {\kern 1pt} {\kern 1pt} {\kern 1pt} {\kern 1pt} {\kern 1pt} {\kern 1pt} {\kern 1pt} {\kern 1pt} {\kern 1pt} {\kern 1pt} {\kern 1pt} {\kern 1pt} {\kern 1pt} {\kern 1pt} {\kern 1pt} {\kern 1pt} {\kern 1pt} {\kern 1pt} {\kern 1pt} {\kern 1pt} {\kern 1pt} {\kern 1pt} {\kern 1pt} {\kern 1pt} {\kern 1pt} {\kern 1pt} {\kern 1pt} {\kern 1pt} {\kern 1pt} {\kern 1pt} {\kern 1pt} {\kern 1pt} {\kern 1pt} {\kern 1pt} {\kern 1pt} {\kern 1pt} {\kern 1pt} {\kern 1pt} \frac{1}{{\sqrt 2 }}\left( {\begin{array}{*{20}{c}}
{ - 1}&{ - 1}\\
1&{ - 1}
\end{array}} \right)\left( {\begin{array}{*{20}{c}}
1\\
1
\end{array}} \right)\\
{\kern 1pt} {\kern 1pt} {\kern 1pt} {\kern 1pt} {\kern 1pt} {\kern 1pt} {\kern 1pt} {\kern 1pt} {\kern 1pt} {\kern 1pt} {\kern 1pt} {\kern 1pt} {\kern 1pt} {\kern 1pt} {\kern 1pt} {\kern 1pt} {\kern 1pt} {\kern 1pt} {\kern 1pt} {\kern 1pt} {\kern 1pt} {\kern 1pt} {\kern 1pt} {\kern 1pt} {\kern 1pt} {\kern 1pt} {\kern 1pt}  = \frac{{\left| 0 \right\rangle  - i\left| 1 \right\rangle }}{{\sqrt 2 }}\\
{\kern 1pt} {\kern 1pt} {\kern 1pt} {\kern 1pt} {\kern 1pt} {\kern 1pt} {\kern 1pt} {\kern 1pt} {\kern 1pt} {\kern 1pt} {\kern 1pt} {\kern 1pt} {\kern 1pt} {\kern 1pt} {\kern 1pt} {\kern 1pt} {\kern 1pt} {\kern 1pt} {\kern 1pt} {\kern 1pt} {\kern 1pt} {\kern 1pt} {\kern 1pt} {\kern 1pt} {\kern 1pt} {\kern 1pt} {\kern 1pt}  = \left| {{y_3}} \right\rangle
\end{array}
\end{array}}
\end{array}
\end{array}
\end{equation}
\textbf{Example 4} (less-complete) Suppose
\begin{equation}
\nonumber
\begin{array}{l}
\left| {{x_1}} \right\rangle  = \frac{{{\rm{a}}\left| 0 \right\rangle  + b\left| 1 \right\rangle }}{{\sqrt {{a^2} + {b^2}} }},\\
\left| {{y_1}} \right\rangle  = \frac{{a\left| 0 \right\rangle  + bi\left| 1 \right\rangle }}{{\sqrt {{a^2} + {b^2}} }}(a = 1,b = 3).
\end{array}
\end{equation}
\textit{Perceptron training}: The quantum perceptron is
\begin{equation}
\nonumber
\begin{array}{*{20}{l}}
{\left| {{y_{output}}} \right\rangle  = \mathop {{\rm{ }}\hat F}{\Sigma _{new}}\mathop {{\hat w_{new}}} \left| {{x_j}} \right\rangle }\\
\begin{array}{l}
{\kern 1pt} {\kern 1pt} {\kern 1pt} {\kern 1pt} {\kern 1pt} {\kern 1pt} {\kern 1pt} {\kern 1pt} {\kern 1pt} {\kern 1pt} {\kern 1pt} {\kern 1pt} {\kern 1pt} {\kern 1pt} {\kern 1pt} {\kern 1pt} {\kern 1pt} {\kern 1pt} {\kern 1pt} {\kern 1pt} {\kern 1pt} {\kern 1pt} {\kern 1pt} {\kern 1pt} {\kern 1pt} {\kern 1pt} {\kern 1pt} {\kern 1pt} {\kern 1pt} {\kern 1pt} {\kern 1pt}  = \frac{1}{{\sqrt {10} }}\left( {\begin{array}{*{20}{c}}
{ - 1}&{3i}\\
{ - 3i}&1
\end{array}} \right)\left( {\begin{array}{*{20}{c}}
1&0\\
0&1
\end{array}} \right) \times \\
{\kern 1pt} {\kern 1pt} {\kern 1pt} {\kern 1pt}{\kern 1pt}{\kern 1pt}{\kern 1pt} {\kern 1pt} {\kern 1pt} {\kern 1pt} {\kern 1pt} {\kern 1pt} {\kern 1pt} {\kern 1pt} {\kern 1pt} {\kern 1pt} {\kern 1pt} {\kern 1pt} {\kern 1pt} {\kern 1pt} {\kern 1pt} {\kern 1pt} {\kern 1pt} {\kern 1pt} {\kern 1pt} {\kern 1pt} {\kern 1pt} {\kern 1pt} {\kern 1pt} {\kern 1pt} {\kern 1pt} {\kern 1pt} {\kern 1pt} {\kern 1pt} {\kern 1pt} {\kern 1pt} {\kern 1pt} {\kern 1pt} {\kern 1pt} {\kern 1pt} {\kern 1pt} {\kern 1pt} {\kern 1pt} {\kern 1pt} \frac{1}{{\sqrt {10} }}\left( {\begin{array}{*{20}{c}}
{ - 1}&{ - 3}\\
{ - 3}&1
\end{array}} \right)\left| {{x_j}} \right\rangle
\end{array}
\end{array}
\end{equation}
\textit{Validation of correctness}:
\begin{equation}
\nonumber
\begin{array}{*{20}{l}}
{\left| {{y_{output}}} \right\rangle  = \mathop {\hat F} {\Sigma _{new}}\mathop {{\hat w_{new}}} \left| {{x_1}} \right\rangle }\\
{\begin{array}{*{20}{l}}
{{\kern 1pt} {\kern 1pt}{\kern 1pt} {\kern 1pt} {\kern 1pt} {\kern 1pt} {\kern 1pt} {\kern 1pt} {\kern 1pt} {\kern 1pt} {\kern 1pt} {\kern 1pt} {\kern 1pt} {\kern 1pt} {\kern 1pt} {\kern 1pt} {\kern 1pt} {\kern 1pt} {\kern 1pt} {\kern 1pt} {\kern 1pt} {\kern 1pt} {\kern 1pt} {\kern 1pt} {\kern 1pt} {\kern 1pt} {\kern 1pt} {\kern 1pt} {\kern 1pt} {\kern 1pt} {\kern 1pt}  = \frac{1}{{\sqrt {10} }}\left( {\begin{array}{*{20}{c}}
{ - 1}&{3i}\\
{ - 3i}&1
\end{array}} \right)\left( {\begin{array}{*{20}{c}}
1&0\\
0&1
\end{array}} \right) \times }\\
{{\kern 1pt} {\kern 1pt} {\kern 1pt}{\kern 1pt} {\kern 1pt}  {\kern 1pt} {\kern 1pt} {\kern 1pt} {\kern 1pt} {\kern 1pt} {\kern 1pt} {\kern 1pt} {\kern 1pt} {\kern 1pt} {\kern 1pt} {\kern 1pt} {\kern 1pt} {\kern 1pt} {\kern 1pt} {\kern 1pt} {\kern 1pt} {\kern 1pt} {\kern 1pt} {\kern 1pt} {\kern 1pt} {\kern 1pt} {\kern 1pt} {\kern 1pt} {\kern 1pt} {\kern 1pt} {\kern 1pt} {\kern 1pt} {\kern 1pt} {\kern 1pt} {\kern 1pt} {\kern 1pt} {\kern 1pt} {\kern 1pt} {\kern 1pt} {\kern 1pt} {\kern 1pt} {\kern 1pt} {\kern 1pt} \frac{1}{{\sqrt {10} }}\left( {\begin{array}{*{20}{c}}
{ - 1}&{ - 3}\\
{ - 3}&1
\end{array}} \right)\frac{1}{{\sqrt {10} }}\left( {\begin{array}{*{20}{l}}
1\\
3
\end{array}} \right)}
\end{array}}\\
\begin{array}{l}
{\kern 1pt}{\kern 1pt} {\kern 1pt}{\kern 1pt} {\kern 1pt} {\kern 1pt} {\kern 1pt} {\kern 1pt} {\kern 1pt} {\kern 1pt} {\kern 1pt} {\kern 1pt} {\kern 1pt} {\kern 1pt} {\kern 1pt} {\kern 1pt} {\kern 1pt} {\kern 1pt} {\kern 1pt} {\kern 1pt} {\kern 1pt} {\kern 1pt} {\kern 1pt} {\kern 1pt} {\kern 1pt} {\kern 1pt} {\kern 1pt} {\kern 1pt} {\kern 1pt}  = \frac{{\left| 0 \right\rangle  + 3i\left| 1 \right\rangle }}{{\sqrt {10} }}\\
{\kern 1pt} {\kern 1pt} {\kern 1pt} {\kern 1pt} {\kern 1pt} {\kern 1pt} {\kern 1pt} {\kern 1pt} {\kern 1pt} {\kern 1pt} {\kern 1pt} {\kern 1pt} {\kern 1pt} {\kern 1pt} {\kern 1pt} {\kern 1pt} {\kern 1pt} {\kern 1pt} {\kern 1pt} {\kern 1pt} {\kern 1pt} {\kern 1pt} {\kern 1pt} {\kern 1pt} {\kern 1pt} {\kern 1pt} {\kern 1pt} {\kern 1pt} {\kern 1pt}  = \left| {{y_1}} \right\rangle
\end{array}
\end{array}
\end{equation}

\subsubsection{$\pi /8$ gate}
\textbf{Example 5} (over-complete) Suppose
\begin{equation}
\nonumber
\begin{array}{*{20}{l}}
{\kern 1pt} {\kern 1pt} {\kern 1pt} {\kern 1pt} {\left| {{x_1}} \right\rangle  = \left| 0 \right\rangle ,\left| {{y_1}} \right\rangle  = \left| 0 \right\rangle }\\
{\kern 1pt} {\kern 1pt} {\kern 1pt} {\kern 1pt} {\left| {{x_2}} \right\rangle  = \left| 1 \right\rangle ,\left| {{y_2}} \right\rangle  = {e^{i\pi /4}}\left| 1 \right\rangle }\\
\begin{array}{l}
\left| {{x_3}} \right\rangle  = \frac{{a\left| 0 \right\rangle  + b\left| 1 \right\rangle }}{{\sqrt {{a^2} + {b^2}} }},\\
\left| {{y_3}} \right\rangle  = \frac{{a\left| 0 \right\rangle  + b{e^{i\pi /4}}\left| 1 \right\rangle }}{{\sqrt {{a^2} + {b^2}} }}(a =  - 8,b =  - 9).
\end{array}
\end{array}
\end{equation}
\textit{Perceptron training}: The quantum perceptron is
\begin{equation}
\nonumber
\begin{array}{*{20}{l}}
{\left| {{y_{output}}} \right\rangle  = \mathop {{\rm{ }}\hat F}  {\Sigma _{new}}\mathop {{\hat w_{new}}}  \left| {{x_j}} \right\rangle }\\
\begin{array}{l}
{\kern 1pt} {\kern 1pt}{\kern 1pt} {\kern 1pt} {\kern 1pt} {\kern 1pt} {\kern 1pt} {\kern 1pt} {\kern 1pt} {\kern 1pt} {\kern 1pt} {\kern 1pt} {\kern 1pt} {\kern 1pt} {\kern 1pt} {\kern 1pt} {\kern 1pt} {\kern 1pt} {\kern 1pt} {\kern 1pt} {\kern 1pt} {\kern 1pt} {\kern 1pt} {\kern 1pt} {\kern 1pt} {\kern 1pt} {\kern 1pt} {\kern 1pt} {\kern 1pt} {\kern 1pt} {\kern 1pt}  = \frac{1}{{\sqrt {145} }}\left( {\begin{array}{*{20}{c}}
{ - 8}&9\\
{ - 9{e^{i\pi /4}}}&{ - 8{e^{i\pi /4}}}
\end{array}} \right)\left( {\begin{array}{*{20}{c}}
1&0\\
0&1
\end{array}} \right) \times \\
{\kern 1pt} {\kern 1pt} {\kern 1pt} {\kern 1pt} {\kern 1pt} {\kern 1pt} {\kern 1pt} {\kern 1pt} {\kern 1pt} {\kern 1pt} {\kern 1pt} {\kern 1pt} {\kern 1pt} {\kern 1pt} {\kern 1pt} {\kern 1pt} {\kern 1pt} {\kern 1pt} {\kern 1pt} {\kern 1pt} {\kern 1pt} {\kern 1pt} {\kern 1pt} {\kern 1pt} {\kern 1pt} {\kern 1pt} {\kern 1pt} {\kern 1pt} {\kern 1pt} {\kern 1pt} {\kern 1pt} {\kern 1pt} {\kern 1pt} {\kern 1pt} {\kern 1pt} {\kern 1pt} {\kern 1pt} {\kern 1pt} {\kern 1pt} {\kern 1pt} {\kern 1pt} {\kern 1pt} {\kern 1pt} {\kern 1pt} \frac{1}{{\sqrt {145} }}\left( {\begin{array}{*{20}{c}}
{ - 8}&{ - 9}\\
9&{ - 8}
\end{array}} \right)\left| {{x_j}} \right\rangle
\end{array}
\end{array}
\end{equation}
\textit{Validation of correctness}:
\begin{equation}
\nonumber
\begin{array}{l}
\left| {{y_{output}}} \right\rangle  = \mathop {\hat F} {\Sigma _{new}}\mathop {{\hat w_{new}}} \left| {{x_1}} \right\rangle  = \left| 0 \right\rangle  = \left| {{y_1}} \right\rangle \\
\left| {{y_{output}}} \right\rangle  = \mathop {\hat F} {\Sigma _{new}}\mathop {{\hat w_{new}}} \left| {{x_2}} \right\rangle  = {e^{i\pi /4}}\left| 1 \right\rangle  = \left| {{y_2}} \right\rangle \\
\left| {{y_{output}}} \right\rangle  = \mathop {\hat F} {\Sigma _{new}}\mathop {{\hat w_{new}}} \left| {{x_3}} \right\rangle  = \frac{{ - 8\left| 0 \right\rangle  - 9{e^{i\pi /4}}\left| 1 \right\rangle }}{{\sqrt {145} }} = \left| {{y_3}} \right\rangle
\end{array}
\end{equation}
\textbf{Example 6} (less-complete) Suppose
\begin{equation}
\nonumber
\begin{array}{l}
\left| {{x_1}} \right\rangle  = \frac{{a\left| 0 \right\rangle  + b\left| 1 \right\rangle }}{{\sqrt {{a^2} + {b^2}} }},\\
\left| {{y_1}} \right\rangle  = \frac{{a\left| 0 \right\rangle  + b{e^{i\pi /4}}\left| 1 \right\rangle }}{{\sqrt {{a^2} + {b^2}} }}(a = 13,b =  - 10).
\end{array}
\end{equation}
\textit{Perceptron training}: The quantum perceptron is
\begin{equation}
\nonumber
\begin{array}{*{20}{l}}
{\left| {{y_{output}}} \right\rangle  = \mathop {{\rm{ }}\hat F} {\Sigma _{new}}\mathop {{\hat w_{new}}} \left| {{x_j}} \right\rangle }\\
\begin{array}{l}
{\kern 1pt} {\kern 1pt} {\kern 1pt} {\kern 1pt} {\kern 1pt} {\kern 1pt} {\kern 1pt} {\kern 1pt} {\kern 1pt} {\kern 1pt} {\kern 1pt} {\kern 1pt} {\kern 1pt} {\kern 1pt} {\kern 1pt} {\kern 1pt} {\kern 1pt} {\kern 1pt} {\kern 1pt} {\kern 1pt} {\kern 1pt} {\kern 1pt} {\kern 1pt} {\kern 1pt} {\kern 1pt} {\kern 1pt} {\kern 1pt} {\kern 1pt} {\kern 1pt} {\kern 1pt} {\kern 1pt}  = \frac{1}{{\sqrt {269} }}\left( {\begin{array}{*{20}{c}}
{ - 13}&{10{e^{i\pi /4}}}\\
{10{e^{i\pi /4}}}&{13i}
\end{array}} \right)\left( {\begin{array}{*{20}{c}}
1&0\\
0&1
\end{array}} \right) \times \\
{\kern 1pt} {\kern 1pt} {\kern 1pt} {\kern 1pt} {\kern 1pt}{\kern 1pt}{\kern 1pt} {\kern 1pt} {\kern 1pt} {\kern 1pt} {\kern 1pt} {\kern 1pt} {\kern 1pt} {\kern 1pt} {\kern 1pt} {\kern 1pt} {\kern 1pt} {\kern 1pt} {\kern 1pt} {\kern 1pt} {\kern 1pt} {\kern 1pt} {\kern 1pt} {\kern 1pt} {\kern 1pt} {\kern 1pt} {\kern 1pt} {\kern 1pt} {\kern 1pt} {\kern 1pt} {\kern 1pt} {\kern 1pt} {\kern 1pt} {\kern 1pt} {\kern 1pt} {\kern 1pt} {\kern 1pt} {\kern 1pt} {\kern 1pt} {\kern 1pt} {\kern 1pt} {\kern 1pt} {\kern 1pt} {\kern 1pt} {\kern 1pt} \frac{1}{{\sqrt {269} }}\left( {\begin{array}{*{20}{c}}
{ - 13}&{10}\\
{ - 10}&{ - 13}
\end{array}} \right)\left| {{x_j}} \right\rangle
\end{array}
\end{array}
\end{equation}
\textit{Validation of correctness:}
\begin{equation}
\nonumber
\begin{array}{l}
\left| {{y_{output}}} \right\rangle  = \mathop {\hat F} {\Sigma _{new}}\mathop {{\hat w_{new}}} \left| {{x_1}} \right\rangle \\
{\kern 1pt} {\kern 1pt} {\kern 1pt} {\kern 1pt} {\kern 1pt} {\kern 1pt} {\kern 1pt} {\kern 1pt} {\kern 1pt} {\kern 1pt} {\kern 1pt} {\kern 1pt} {\kern 1pt} {\kern 1pt} {\kern 1pt} {\kern 1pt} {\kern 1pt} {\kern 1pt} {\kern 1pt} {\kern 1pt} {\kern 1pt} {\kern 1pt} {\kern 1pt} {\kern 1pt} {\kern 1pt} {\kern 1pt} {\kern 1pt} {\kern 1pt} {\kern 1pt} {\kern 1pt} {\kern 1pt} {\kern 1pt} {\kern 1pt} {\kern 1pt} {\kern 1pt}  = \frac{{13\left| 0 \right\rangle  - 10{e^{i\pi /4}}\left| 1 \right\rangle }}{{\sqrt {269} }} = \left| {{y_1}} \right\rangle
\end{array}
\end{equation}

\subsubsection{\textit{CNOT} gate}
\textbf{Example 7} (over-complete) Suppose that the example is:
\begin{equation}
\nonumber
\begin{array}{l}
\left| {{x_1}} \right\rangle  = \left| {00} \right\rangle ,{\kern 1pt} {\kern 1pt} {\kern 1pt} {\kern 1pt} {\kern 1pt} {\kern 1pt} {\kern 1pt} \left| {{y_1}} \right\rangle  = \left| {00} \right\rangle \\
\left| {{x_2}} \right\rangle  = \left| {01} \right\rangle ,{\kern 1pt} {\kern 1pt} {\kern 1pt} {\kern 1pt} {\kern 1pt} {\kern 1pt} {\kern 1pt} \left| {{y_2}} \right\rangle  = \left| {01} \right\rangle \\
\left| {{x_3}} \right\rangle  = \left| {10} \right\rangle ,{\kern 1pt} {\kern 1pt} {\kern 1pt} {\kern 1pt} {\kern 1pt} {\kern 1pt} {\kern 1pt} \left| {{y_3}} \right\rangle  = \left| {11} \right\rangle \\
\left| {{x_4}} \right\rangle  = \left| {11} \right\rangle ,{\kern 1pt} {\kern 1pt} {\kern 1pt} {\kern 1pt} {\kern 1pt} {\kern 1pt} {\kern 1pt} \left| {{y_4}} \right\rangle  = \left| {10} \right\rangle \\
\left| {{x_5}} \right\rangle  = \frac{{a\left| {00} \right\rangle  + b\left| {11} \right\rangle }}{{\sqrt {{a^2} + {b^2}} }},{\kern 1pt} {\kern 1pt} {\kern 1pt} {\kern 1pt} {\kern 1pt} {\kern 1pt} {\kern 1pt} \left| {{y_5}} \right\rangle  = \frac{{a\left| {00} \right\rangle  + b\left| {10} \right\rangle }}{{\sqrt {{a^2} + {b^2}} }}(a = 3,b = 4)
\end{array}
\end{equation}
\textit{Perceptron training:} The same way through Algorithm~\ref{alg:2} for training and learning and finally get the quantum perceptron:
\begin{equation}
\nonumber
\begin{array}{*{20}{l}}
{\left| {{y_{output}}} \right\rangle  = \mathop {{\rm{ }}\hat F} {\Sigma _{new}}\mathop {{\hat w_{new}}} \left| {{x_j}} \right\rangle }\\
\begin{array}{l}
{\kern 1pt} {\kern 1pt} {\kern 1pt} {\kern 1pt}{\kern 1pt} {\kern 1pt} {\kern 1pt} {\kern 1pt} {\kern 1pt} {\kern 1pt} {\kern 1pt} {\kern 1pt} {\kern 1pt} {\kern 1pt} {\kern 1pt} {\kern 1pt} {\kern 1pt} {\kern 1pt} {\kern 1pt} {\kern 1pt} {\kern 1pt} {\kern 1pt} {\kern 1pt} {\kern 1pt} {\kern 1pt} {\kern 1pt} {\kern 1pt} {\kern 1pt} {\kern 1pt}  = \left( {\begin{array}{*{20}{c}}
{ - 0.6}&{0.8}&0&0\\
0&0&0&{ - 1}\\
{ - 0.8}&{ - 0.6}&0&0\\
0&0&1&0
\end{array}} \right)\left( {\begin{array}{*{20}{c}}
1&0&0&0\\
0&1&0&0\\
0&0&1&0\\
0&0&0&1
\end{array}} \right) \times \\
{\kern 1pt} {\kern 1pt} {\kern 1pt} {\kern 1pt} {\kern 1pt} {\kern 1pt} {\kern 1pt} {\kern 1pt} {\kern 1pt} {\kern 1pt} {\kern 1pt} {\kern 1pt} {\kern 1pt} {\kern 1pt} {\kern 1pt} {\kern 1pt} {\kern 1pt} {\kern 1pt} {\kern 1pt} {\kern 1pt} {\kern 1pt} {\kern 1pt} {\kern 1pt} {\kern 1pt} {\kern 1pt} {\kern 1pt} {\kern 1pt} {\kern 1pt} {\kern 1pt} {\kern 1pt} {\kern 1pt} {\kern 1pt} {\kern 1pt} {\kern 1pt} {\kern 1pt} {\kern 1pt} {\kern 1pt} {\kern 1pt} {\kern 1pt} {\kern 1pt} {\kern 1pt} {\kern 1pt} \left( {\begin{array}{*{20}{c}}
{ - 0.6}&0&0&{ - 0.8}\\
{0.8}&0&0&{ - 0.6}\\
0&0&1&0\\
0&{ - 1}&0&0
\end{array}} \right)\left| {{x_j}} \right\rangle
\end{array}
\end{array}
\end{equation}
\textit{Validation of correctness:}
\begin{equation}
\nonumber
\begin{array}{l}
\left| {{y_{output}}} \right\rangle  = \mathop {\hat F} {\Sigma _{new}}\mathop {{\hat w_{new}}} \left| {{x_1}} \right\rangle  = \left| {00} \right\rangle  = \left| {{y_1}} \right\rangle \\
\left| {{y_{output}}} \right\rangle  = \mathop {\hat F} {\Sigma _{new}}\mathop {{\hat w_{new}}} \left| {{x_2}} \right\rangle  = \left| {01} \right\rangle  = \left| {{y_2}} \right\rangle \\
\left| {{y_{output}}} \right\rangle  = \mathop {\hat F} {\Sigma _{new}}\mathop {{\hat w_{new}}} \left| {{x_3}} \right\rangle  = \left| {11} \right\rangle  = \left| {{y_3}} \right\rangle \\
\left| {{y_{output}}} \right\rangle  = \mathop {\hat F} {\Sigma _{new}}\mathop {{\hat w_{new}}} \left| {{x_4}} \right\rangle  = \left| {10} \right\rangle  = \left| {{y_4}} \right\rangle \\
\left| {{y_{output}}} \right\rangle  = \mathop {\hat F} {\Sigma _{new}}\mathop {{\hat w_{new}}} \left| {{x_5}} \right\rangle  = \frac{{3\left| {00} \right\rangle  + 4\left| {10} \right\rangle }}{5} = \left| {{y_5}} \right\rangle
\end{array}
\end{equation}
\textbf{Example 8} (less-complete) Suppose
\begin{equation}
\nonumber
\begin{array}{l}
\left| {{x_1}} \right\rangle  = \left| {00} \right\rangle ,{\kern 1pt} {\kern 1pt} {\kern 1pt} {\kern 1pt} {\kern 1pt} {\kern 1pt} {\kern 1pt} \left| {{y_1}} \right\rangle  = \left| {00} \right\rangle \\
\left| {{x_2}} \right\rangle  = \frac{{a\left| {01} \right\rangle  + b\left| {11} \right\rangle }}{{\sqrt {{a^2} + {b^2}} }},{\kern 1pt} {\kern 1pt} {\kern 1pt} {\kern 1pt} {\kern 1pt} {\kern 1pt} {\kern 1pt} \left| {{y_2}} \right\rangle  = \frac{{a\left| {01} \right\rangle  + b\left| {10} \right\rangle }}{{\sqrt {{a^2} + {b^2}} }}(a = 5,b =  - 6)
\end{array}
\end{equation}
\textit{Perceptron training}: The quantum perceptron is
\begin{equation}
\nonumber
\begin{array}{*{20}{l}}
{\left| {{y_{output}}} \right\rangle  = \mathop {\hat F} {\Sigma _{new}}\mathop {{\hat w_{new}}} \left| {{x_j}} \right\rangle \sum\limits_{i = 1}^n {{X_i}} }\\
{\begin{array}{*{20}{l}}
{{\kern 1pt} {\kern 1pt} {\kern 1pt} {\kern 1pt} {\kern 1pt} {\kern 1pt} {\kern 1pt} {\kern 1pt}  = \frac{1}{{\sqrt {61} }}\left( {\begin{array}{*{20}{c}}
0&{\sqrt {61} }&0&0\\
{ - 5}&0&0&6\\
6&0&0&5\\
0&0&{\sqrt {61} }&0
\end{array}} \right)\left( {\begin{array}{*{20}{c}}
1&0&0&0\\
0&1&0&0\\
0&0&1&0\\
0&0&0&1
\end{array}} \right) \times }\\
{{\kern 1pt} {\kern 1pt} {\kern 1pt} {\kern 1pt}{\kern 1pt} {\kern 1pt}{\kern 1pt} {\kern 1pt}{\kern 1pt} {\kern 1pt}{\kern 1pt} {\kern 1pt}{\kern 1pt} {\kern 1pt}{\kern 1pt} {\kern 1pt} {\kern 1pt} {\kern 1pt} {\kern 1pt} {\kern 1pt} \frac{1}{{\sqrt {61} }}\left( {\begin{array}{*{20}{c}}
0&{ - 5}&6&0\\
{\sqrt {61} }&0&0&0\\
0&0&0&{\sqrt {61} }\\
0&6&5&0
\end{array}} \right)\left| {{x_j}} \right\rangle }
\end{array}}
\end{array}
\end{equation}
\textit{Validation of correctness}:
\begin{equation}
\nonumber
\begin{array}{l}
\left| {{y_{output}}} \right\rangle  = \mathop {\hat F} {\Sigma _{new}}\mathop {{\hat w_{new}}} \left| {{x_1}} \right\rangle  = \left| {00} \right\rangle  = \left| {{y_1}} \right\rangle \\
\left| {{y_{output}}} \right\rangle  = \mathop {\hat F} {\Sigma _{new}}\mathop {{\hat w_{new}}} \left| {{x_2}} \right\rangle  = \frac{{5\left| {01} \right\rangle  - 6\left| {10} \right\rangle }}{{\sqrt {61} }} = \left| {{y_2}} \right\rangle
\end{array}
\end{equation}

\subsection{Complex quantum gates}
In addition to the basic quantum gates illustrated before, there are also some complex gates which are frequently used, such as the \textit{Toffoli} and \textit{Fredkin} gates. Here, we take these two gates as an example to validate the applicability of our proposed algorithm.

\subsubsection{\textit{Toffoli} gate}
\begin{figure}
\centering
\includegraphics[height=2cm]{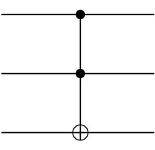}
\caption{Quantum circuit of the \textit{Toffoli} gate.}
\label{fig:3}
\end{figure}

The \textit{Toffoli} gate has three input bits and three output bits. Two of the bits are control bits that are unaffected by the action of the \textit{Toffoli} gate. The third bit is a target bit that is flipped if both control bits are set to 1, and otherwise is left alone. The Fig.~\ref{fig:3} illustrates the circuit of \textit{Toffoli} gate, and Eq.~\ref{12} gives the matrix form of \textit{Toffoli} operator.

\begin{equation}
{\rm{\textit{Toffoli}}} =
{\begin{pmatrix}
    1&0&0&0&0&0&0&0\\
    0&1&0&0&0&0&0&0\\
    0&0&1&0&0&0&0&0\\
    0&0&0&1&0&0&0&0\\
    0&0&0&0&1&0&0&0\\
    0&0&0&0&0&1&0&0\\
    0&0&0&0&0&0&0&1\\
    0&0&0&0&0&0&1&0
\end{pmatrix}}
\label{12}
\end{equation}
\textbf{Example 9} (over-complete) Suppose
\begin{equation}
\nonumber
\begin{array}{l}
\left| {{x_1}} \right\rangle  = \left| {000} \right\rangle ,{\kern 1pt} {\kern 1pt} {\kern 1pt} {\kern 1pt} {\kern 1pt} {\kern 1pt} {\kern 1pt} \left| {{y_1}} \right\rangle  = \left| {000} \right\rangle .{\kern 1pt} {\kern 1pt} {\kern 1pt} {\kern 1pt} \left| {{x_2}} \right\rangle  = \left| {001} \right\rangle ,{\kern 1pt} {\kern 1pt} {\kern 1pt} {\kern 1pt} {\kern 1pt} {\kern 1pt} {\kern 1pt} \left| {{y_2}} \right\rangle  = \left| {001} \right\rangle \\
\left| {{x_3}} \right\rangle  = \left| {010} \right\rangle ,{\kern 1pt} {\kern 1pt} {\kern 1pt} {\kern 1pt} {\kern 1pt} {\kern 1pt} {\kern 1pt} \left| {{y_3}} \right\rangle  = \left| {010} \right\rangle .{\kern 1pt} {\kern 1pt} {\kern 1pt} {\kern 1pt} \left| {{x_4}} \right\rangle  = \left| {011} \right\rangle ,{\kern 1pt} {\kern 1pt} {\kern 1pt} {\kern 1pt} {\kern 1pt} {\kern 1pt} {\kern 1pt} \left| {{y_4}} \right\rangle  = \left| {011} \right\rangle \\
\left| {{x_5}} \right\rangle  = \left| {100} \right\rangle ,{\kern 1pt} {\kern 1pt} {\kern 1pt} {\kern 1pt} {\kern 1pt} {\kern 1pt} {\kern 1pt} \left| {{y_5}} \right\rangle  = \left| {100} \right\rangle .{\kern 1pt} {\kern 1pt} {\kern 1pt} {\kern 1pt} \left| {{x_6}} \right\rangle  = \left| {101} \right\rangle ,{\kern 1pt} {\kern 1pt} {\kern 1pt} {\kern 1pt} {\kern 1pt} {\kern 1pt} {\kern 1pt} \left| {{y_6}} \right\rangle  = \left| {101} \right\rangle \\
\left| {{x_7}} \right\rangle  = \left| {110} \right\rangle ,{\kern 1pt} {\kern 1pt} {\kern 1pt} {\kern 1pt} {\kern 1pt} {\kern 1pt} {\kern 1pt} \left| {{y_7}} \right\rangle  = \left| {111} \right\rangle .{\kern 1pt} {\kern 1pt} {\kern 1pt} {\kern 1pt} \left| {{x_8}} \right\rangle  = \left| {111} \right\rangle ,{\kern 1pt} {\kern 1pt} {\kern 1pt} {\kern 1pt} {\kern 1pt} {\kern 1pt} {\kern 1pt} \left| {{y_8}} \right\rangle  = \left| {110} \right\rangle \\
\left| {{x_9}} \right\rangle  = \frac{{a\left| {001} \right\rangle  + b\left| {110} \right\rangle }}{{\sqrt {{a^2} + {b^2}} }},{\kern 1pt} {\kern 1pt} {\kern 1pt} {\kern 1pt} {\kern 1pt} {\kern 1pt} {\kern 1pt} \left| {{y_9}} \right\rangle  = \frac{{a\left| {001} \right\rangle  + b\left| {111} \right\rangle }}{{\sqrt {{a^2} + {b^2}} }}(a = 3,b = 4)
\end{array}
\end{equation}
\textit{Perceptron training}: The quantum perceptron is
\begin{equation}
\nonumber
\begin{array}{*{20}{l}}
{\left| {{y_{output}}} \right\rangle  = \mathop {{\rm{ }}\hat F} {\Sigma _{new}}\mathop {{\hat w_{new}}}  \left| {{x_j}} \right\rangle }\\
\begin{array}{l}
{\kern 1pt} {\kern 1pt} {\kern 1pt} {\kern 1pt} {\kern 1pt} {\kern 1pt} {\kern 1pt} {\kern 1pt} {\kern 1pt} {\kern 1pt} {\kern 1pt} {\kern 1pt} {\kern 1pt}  = \frac{1}{5}\left( {\begin{smallmatrix}
0&0&0&0&0&0&0&5\\
{ - 3}&4&0&0&0&0&0&0\\
0&0&0&0&0&0&{ - 5}&0\\
0&0&0&0&{ - 5}&0&0&0\\
0&0&5&0&0&0&0&0\\
0&0&0&5&0&0&0&0\\
0&0&0&0&0&{ - 5}&0&0\\
{ - 4}&{ - 3}&0&0&0&0&0&0
\end{smallmatrix}} \right) \times \\
{\kern 1pt} {\kern 1pt} {\kern 1pt} {\kern 1pt}{\kern 1pt} {\kern 1pt}{\kern 1pt} {\kern 1pt}{\kern 1pt} {\kern 1pt} {\kern 1pt} {\kern 1pt} {\kern 1pt} {\kern 1pt} {\kern 1pt} {\kern 1pt} {\kern 1pt} {\kern 1pt} {\kern 1pt} {\kern 1pt} {\kern 1pt} {\kern 1pt} {\kern 1pt} {\kern 1pt} {\kern 1pt} {\kern 1pt} {\kern 1pt} {\kern 1pt} {\kern 1pt} {\kern 1pt} {\kern 1pt} {\kern 1pt} {\kern 1pt} {\kern 1pt} \left( {\begin{smallmatrix}
1&0&0&0&0&0&0&0\\
0&1&0&0&0&0&0&0\\
0&0&1&0&0&0&0&0\\
0&0&0&1&0&0&0&0\\
0&0&0&0&1&0&0&0\\
0&0&0&0&0&1&0&0\\
0&0&0&0&0&0&1&0\\
0&0&0&0&0&0&0&1
\end{smallmatrix}} \right) \times \\
{\kern 1pt} {\kern 1pt} {\kern 1pt} {\kern 1pt} {\kern 1pt} {\kern 1pt} {\kern 1pt} {\kern 1pt} {\kern 1pt} {\kern 1pt} {\kern 1pt} {\kern 1pt} {\kern 1pt} {\kern 1pt} {\kern 1pt} {\kern 1pt} {\kern 1pt} {\kern 1pt} {\kern 1pt} {\kern 1pt} {\kern 1pt} {\kern 1pt} \frac{1}{5}\left( {\begin{smallmatrix}
0&{ - 3}&0&0&0&0&{ - 4}&0\\
0&4&0&0&0&0&{ - 3}&0\\
0&0&0&0&5&0&0&0\\
0&0&0&0&0&5&0&0\\
0&0&0&{ - 5}&0&0&0&0\\
0&0&0&0&0&0&0&{ - 5}\\
0&0&{ - 5}&0&0&0&0&0\\
5&0&0&0&0&0&0&0
\end{smallmatrix}} \right)\left| {{x_j}} \right\rangle
\end{array}
\end{array}
\end{equation}
\textit{Validation of correctness}:
\begin{equation}
\nonumber
\begin{array}{l}
\left| {{y_{output}}} \right\rangle  = \mathop {\hat F} {\Sigma _{new}}\mathop {{\hat w_{new}}} \left| {{x_1}} \right\rangle  = \left| {000} \right\rangle  = \left| {{y_1}} \right\rangle \\
\left| {{y_{output}}} \right\rangle  = \mathop {\hat F} {\Sigma _{new}}\mathop {{\hat w_{new}}}  \left| {{x_2}} \right\rangle  = \left| {001} \right\rangle  = \left| {{y_2}} \right\rangle \\
\left| {{y_{output}}} \right\rangle  = \mathop {\hat F} {\Sigma _{new}}\mathop {{\hat w_{new}}} \left| {{x_3}} \right\rangle  = \left| {010} \right\rangle  = \left| {{y_3}} \right\rangle \\
\left| {{y_{output}}} \right\rangle  = \mathop {\hat F} {\Sigma _{new}}\mathop {{\hat w_{new}}}  {\kern 1pt} \left| {{x_4}} \right\rangle  = \left| {011} \right\rangle  = {\kern 1pt} \left| {{y_4}} \right\rangle \\
\left| {{y_{output}}} \right\rangle  = \mathop {\hat F} {\Sigma _{new}}\mathop {{\hat w_{new}}} {\kern 1pt} \left| {{x_5}} \right\rangle  = \left| {100} \right\rangle  = \left| {{y_5}} \right\rangle \\
\left| {{y_{output}}} \right\rangle  = \mathop {\hat F} {\Sigma _{new}}\mathop {{\hat w_{new}}} {\kern 1pt} \left| {{x_6}} \right\rangle  = \left| {101} \right\rangle  = \left| {{y_6}} \right\rangle \\
\left| {{y_{output}}} \right\rangle  = \mathop {\hat F} {\Sigma _{new}}\mathop {{\hat w_{new}}} \left| {{x_7}} \right\rangle  = \left| {111} \right\rangle  = \left| {{y_7}} \right\rangle \\
\left| {{y_{output}}} \right\rangle  = \mathop {\hat F} {\Sigma _{new}}\mathop {{\hat w_{new}}} \left| {{x_8}} \right\rangle  = \left| {110} \right\rangle  = \left| {{y_8}} \right\rangle \\
\left| {{y_{output}}} \right\rangle  = \mathop {\hat F} {\Sigma _{new}}\mathop {{\hat w_{new}}} \left| {{x_9}} \right\rangle  = \frac{{3\left| {001} \right\rangle  + 3\left| {111} \right\rangle }}{{\sqrt {{3^2} + {4^2}} }} = \left| {{y_9}} \right\rangle
\end{array}
\end{equation}
\textbf{Example 10} (less-complete) Suppose
\begin{equation}
\nonumber
\begin{array}{l}
\left| {{x_1}} \right\rangle  = \left| {000} \right\rangle ,{\kern 1pt} {\kern 1pt} {\kern 1pt} {\kern 1pt} {\kern 1pt} {\kern 1pt} {\kern 1pt} \left| {{y_1}} \right\rangle  = \left| {000} \right\rangle \\
\left| {{x_2}} \right\rangle  = \left| {110} \right\rangle ,{\kern 1pt} {\kern 1pt} {\kern 1pt} {\kern 1pt} {\kern 1pt} {\kern 1pt} {\kern 1pt} \left| {{y_2}} \right\rangle  = \left| {111} \right\rangle \\
\left| {{x_3}} \right\rangle  = \frac{{a\left| {101} \right\rangle  + b\left| {111} \right\rangle }}{{\sqrt {{a^2} + {b^2}} }},{\kern 1pt} {\kern 1pt} {\kern 1pt} {\kern 1pt} {\kern 1pt} {\kern 1pt} {\kern 1pt} \left| {{y_3}} \right\rangle  = \frac{{a\left| {101} \right\rangle  + b\left| {110} \right\rangle }}{{\sqrt {{a^2} + {b^2}} }}(a = 1,b = 2)
\end{array}
\end{equation}
\textit{Perceptron training}: The quantum perceptron is
\begin{equation}
\nonumber
\begin{array}{*{20}{l}}
{\left| {{y_{output}}} \right\rangle  = \mathop {{\rm{ }}\hat F} {\Sigma _{new}}\mathop {{\hat w_{new}}} \left| {{x_j}} \right\rangle }\\
\begin{array}{l}
{\kern 1pt} {\kern 1pt} {\kern 1pt} {\kern 1pt} {\kern 1pt} {\kern 1pt} {\kern 1pt} {\kern 1pt} {\kern 1pt} {\kern 1pt} {\kern 1pt} {\kern 1pt} {\kern 1pt}  = \frac{1}{{\sqrt 5 }}\left( {\begin{smallmatrix}
0&{\sqrt 5 }&0&0&0&0&0&0\\
0&0&0&0&0&0&0&{\sqrt 5 }\\
0&0&0&0&0&{\sqrt 5 }&0&0\\
0&0&0&0&0&0&{\sqrt 5 }&0\\
0&0&0&0&{\sqrt 5 }&0&0&0\\
{ - 1}&0&0&{ - 2}&0&0&0&0\\
{ - 2}&0&0&1&0&0&0&0\\
0&0&{\sqrt 5 }&0&0&0&0&0
\end{smallmatrix}} \right) \times \\
{\kern 1pt} {\kern 1pt} {\kern 1pt} {\kern 1pt} {\kern 1pt} {\kern 1pt} {\kern 1pt}{\kern 1pt} {\kern 1pt}{\kern 1pt} {\kern 1pt}{\kern 1pt} {\kern 1pt}{\kern 1pt} {\kern 1pt}{\kern 1pt} {\kern 1pt}{\kern 1pt} {\kern 1pt} {\kern 1pt} {\kern 1pt} {\kern 1pt} {\kern 1pt} {\kern 1pt} {\kern 1pt} {\kern 1pt} {\kern 1pt} {\kern 1pt} {\kern 1pt} {\kern 1pt} {\kern 1pt} {\kern 1pt} {\kern 1pt} {\kern 1pt} {\kern 1pt} {\kern 1pt} {\kern 1pt} {\kern 1pt} \left( {\begin{smallmatrix}
1&0&0&0&0&0&0&0\\
0&1&0&0&0&0&0&0\\
0&0&1&0&0&0&0&0\\
0&0&0&1&0&0&0&0\\
0&0&0&0&1&0&0&0\\
0&0&0&0&0&1&0&0\\
0&0&0&0&0&0&1&0\\
0&0&0&0&0&0&0&1
\end{smallmatrix}} \right) \times \\
{\kern 1pt} {\kern 1pt} {\kern 1pt} {\kern 1pt} {\kern 1pt} {\kern 1pt}{\kern 1pt} {\kern 1pt}{\kern 1pt} {\kern 1pt} {\kern 1pt} {\kern 1pt} {\kern 1pt} {\kern 1pt} {\kern 1pt} {\kern 1pt} {\kern 1pt} {\kern 1pt} {\kern 1pt} {\kern 1pt} {\kern 1pt} {\kern 1pt} \frac{1}{{\sqrt 5 }}\left( {\begin{smallmatrix}
0&0&0&0&0&{ - 1}&0&{ - 2}\\
{\sqrt 5 }&0&0&0&0&0&0&0\\
0&0&0&0&0&0&{\sqrt 5 }&0\\
0&0&0&0&0&{ - 2}&0&1\\
0&0&0&0&{\sqrt 5 }&0&0&0\\
0&0&{\sqrt 5 }&0&0&0&0&0\\
0&0&0&{\sqrt 5 }&0&0&0&0\\
0&{\sqrt 5 }&0&0&0&0&0&0
\end{smallmatrix}} \right)\left| {{x_j}} \right\rangle
\end{array}
\end{array}
\end{equation}
\textit{Validation of correctness}:
\begin{equation}
\nonumber
\begin{array}{l}
\left| {{y_{output}}} \right\rangle  = \mathop {\hat F} {\Sigma _{new}}\mathop {{\hat w_{new}}} \left| {{x_1}} \right\rangle  = \left| {000} \right\rangle  = \left| {{y_1}} \right\rangle \\
\left| {{y_{output}}} \right\rangle  = \mathop {\hat F} {\Sigma _{new}}\mathop {{\hat w_{new}}} \left| {{x_2}} \right\rangle  = \left| {111} \right\rangle  = \left| {{y_2}} \right\rangle \\
\left| {{y_{output}}} \right\rangle  = \mathop {\hat F} {\Sigma _{new}}\mathop {{\hat w_{new}}} \left| {{x_3}} \right\rangle  = \frac{{\left| {101} \right\rangle  + 2\left| {110} \right\rangle }}{{\sqrt 5 }} = \left| {{y_3}} \right\rangle
\end{array}
\end{equation}

\subsubsection{$Fredkin$ gate}
\begin{figure}
\centering
\includegraphics[height=2cm]{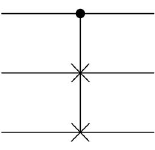}
\caption{Quantum circuit of the \textit{Fredkin} gate.}
\label{fig:4}
\end{figure}

The \textit{Fredkin} gate (also called \textit{CSWAP} gate) is a computational circuit suitable for reversible computation. The \textit{Fredkin} gate is a circuit or device with three inputs and three outputs, the first bit is control bit and the remaining two bits are target bits. The Fig.~\ref{fig:4} shows that the \textit{Fredkin} gate can transmit the first bit unchanged and swap the last two bits if the first bit is 1, and the Eq.~\ref{13} gives the matrix form of \textit{Fredkin} operator.

\begin{equation}
{\rm{\textit{Fredkin}}} = \left(
{\begin{array}{*{20}{c}}
   1 & 0 & 0 & 0 & 0 & 0 & 0 & 0  \cr
   0 & 1 & 0 & 0 & 0 & 0 & 0 & 0  \cr
   0 & 0 & 1 & 0 & 0 & 0 & 0 & 0  \cr
   0 & 0 & 0 & 1 & 0 & 0 & 0 & 0  \cr
   0 & 0 & 0 & 0 & 1 & 0 & 0 & 0  \cr
   0 & 0 & 0 & 0 & 0 & 0 & 1 & 0  \cr
   0 & 0 & 0 & 0 & 0 & 1 & 0 & 0  \cr
   0 & 0 & 0 & 0 & 0 & 0 & 0 & 1  \cr
\end{array}} \right)
\label{13}
\end{equation}

\noindent\textbf{Example 11} (over-complete) Suppose
\begin{equation}
\nonumber
\begin{array}{l}
\left| {{x_1}} \right\rangle  = \left| {000} \right\rangle ,{\kern 1pt} {\kern 1pt} {\kern 1pt} {\kern 1pt} {\kern 1pt} {\kern 1pt} {\kern 1pt} \left| {{y_1}} \right\rangle  = \left| {000} \right\rangle .{\kern 1pt} {\kern 1pt} {\kern 1pt} {\kern 1pt} \left| {{x_2}} \right\rangle  = \left| {001} \right\rangle ,{\kern 1pt} {\kern 1pt} {\kern 1pt} {\kern 1pt} {\kern 1pt} {\kern 1pt} {\kern 1pt} \left| {{y_2}} \right\rangle  = \left| {001} \right\rangle \\
\left| {{x_3}} \right\rangle  = \left| {010} \right\rangle ,{\kern 1pt} {\kern 1pt} {\kern 1pt} {\kern 1pt} {\kern 1pt} {\kern 1pt} {\kern 1pt} \left| {{y_3}} \right\rangle  = \left| {010} \right\rangle .{\kern 1pt} {\kern 1pt} {\kern 1pt} {\kern 1pt} \left| {{x_4}} \right\rangle  = \left| {011} \right\rangle ,{\kern 1pt} {\kern 1pt} {\kern 1pt} {\kern 1pt} {\kern 1pt} {\kern 1pt} {\kern 1pt} \left| {{y_4}} \right\rangle  = \left| {011} \right\rangle \\
\left| {{x_5}} \right\rangle  = \left| {100} \right\rangle ,{\kern 1pt} {\kern 1pt} {\kern 1pt} {\kern 1pt} {\kern 1pt} {\kern 1pt} {\kern 1pt} \left| {{y_5}} \right\rangle  = \left| {100} \right\rangle .{\kern 1pt} {\kern 1pt} {\kern 1pt} {\kern 1pt} \left| {{x_6}} \right\rangle  = \left| {101} \right\rangle ,{\kern 1pt} {\kern 1pt} {\kern 1pt} {\kern 1pt} {\kern 1pt} {\kern 1pt} {\kern 1pt} \left| {{y_6}} \right\rangle  = \left| {110} \right\rangle \\
\left| {{x_7}} \right\rangle  = \left| {110} \right\rangle ,{\kern 1pt} {\kern 1pt} {\kern 1pt} {\kern 1pt} {\kern 1pt} {\kern 1pt} {\kern 1pt} \left| {{y_7}} \right\rangle  = \left| {101} \right\rangle .{\kern 1pt} {\kern 1pt} {\kern 1pt} {\kern 1pt} \left| {{x_8}} \right\rangle  = \left| {111} \right\rangle ,{\kern 1pt} {\kern 1pt} {\kern 1pt} {\kern 1pt} {\kern 1pt} {\kern 1pt} {\kern 1pt} \left| {{y_8}} \right\rangle  = \left| {111} \right\rangle \\
\left| {{x_9}} \right\rangle  = \frac{{a\left| {010} \right\rangle  + b\left| {101} \right\rangle }}{{\sqrt {{a^2} + {b^2}} }},{\kern 1pt} {\kern 1pt} {\kern 1pt} {\kern 1pt} {\kern 1pt} {\kern 1pt} {\kern 1pt} \left| {{y_9}} \right\rangle  = \frac{{a\left| {010} \right\rangle  + b\left| {110} \right\rangle }}{{\sqrt {{a^2} + {b^2}} }}(a = 2,b =  - 7)
\end{array}
\end{equation}
\textit{Perceptron training}: The quantum perceptron is
\begin{equation}
\nonumber
\begin{array}{*{20}{l}}
{\left| {{y_{output}}} \right\rangle  = \mathop {{\rm{ }}\hat F}  {\Sigma _{new}}\mathop {{\hat w_{new}}}  \left| {{x_j}} \right\rangle }\\
\begin{array}{l}
 {\kern 1pt} {\kern 1pt} {\kern 1pt} {\kern 1pt} {\kern 1pt}  = \frac{1}{{\sqrt {53} }}\left( {\begin{smallmatrix}
0&0&0&0&0&0&0&{\sqrt {53} }\\
0&0&0&0&0&0&{\sqrt {53} }&0\\
{ - 2}&7&0&0&0&0&0&0\\
0&0&0&{\sqrt {53} }&0&0&0&0\\
0&0&{\sqrt {53} }&0&0&0&0&0\\
0&0&0&0&{\sqrt {53} }&0&0&0\\
7&2&0&0&0&0&0&0\\
0&0&0&0&0&{\sqrt {53} }&0&0
\end{smallmatrix}} \right) \times \\
{\kern 1pt} {\kern 1pt} {\kern 1pt} {\kern 1pt}{\kern 1pt}{\kern 1pt} {\kern 1pt}{\kern 1pt} {\kern 1pt}{\kern 1pt} {\kern 1pt}{\kern 1pt} {\kern 1pt}{\kern 1pt} {\kern 1pt}{\kern 1pt} {\kern 1pt} {\kern 1pt}{\kern 1pt} {\kern 1pt}{\kern 1pt} {\kern 1pt}{\kern 1pt} {\kern 1pt} {\kern 1pt} {\kern 1pt} {\kern 1pt} {\kern 1pt} {\kern 1pt} {\kern 1pt} {\kern 1pt} {\kern 1pt} \left( {\begin{smallmatrix}
1&0&0&0&0&0&0&0\\
0&1&0&0&0&0&0&0\\
0&0&1&0&0&0&0&0\\
0&0&0&1&0&0&0&0\\
0&0&0&0&1&0&0&0\\
0&0&0&0&0&1&0&0\\
0&0&0&0&0&0&1&0\\
0&0&0&0&0&0&0&1
\end{smallmatrix}} \right) \times \\
{\kern 1pt} {\kern 1pt} {\kern 1pt} {\kern 1pt} {\kern 1pt}{\kern 1pt} {\kern 1pt}{\kern 1pt} {\kern 1pt}{\kern 1pt} {\kern 1pt}{\kern 1pt} {\kern 1pt}{\kern 1pt} \frac{1}{{\sqrt {53} }}\left( {\begin{smallmatrix}
0&0&{ - 2}&0&0&7&0&0\\
0&0&7&0&0&2&0&0\\
0&0&0&0&{\sqrt {53} }&0&0&0\\
0&0&0&{\sqrt {53} }&0&0&0&0\\
0&0&0&0&0&0&{\sqrt {53} }&0\\
0&0&0&0&0&0&0&{\sqrt {53} }\\
0&{\sqrt {53} }&0&0&0&0&0&0\\
{\sqrt {53} }&0&0&0&0&0&0&0
\end{smallmatrix}} \right)\left| {{x_j}} \right\rangle
\end{array}
\end{array}
\end{equation}
\textit{Validation of correctness}:
\begin{equation}
\nonumber
\begin{array}{l}
\left| {{y_{output}}} \right\rangle  = \mathop {\hat F}{\Sigma _{new}}\mathop {{\hat w_{new}}} \left| {{x_1}} \right\rangle  = \left| {000} \right\rangle  = \left| {{y_1}} \right\rangle \\
\left| {{y_{output}}} \right\rangle  = \mathop {\hat F}{\Sigma _{new}}\mathop {{\hat w_{new}}} \left| {{x_2}} \right\rangle  = \left| {001} \right\rangle  = \left| {{y_2}} \right\rangle \\
\left| {{y_{output}}} \right\rangle  = \mathop {\hat F} {\Sigma _{new}}\mathop {{\hat w_{new}}} \left| {{x_3}} \right\rangle  = \left| {010} \right\rangle  = \left| {{y_3}} \right\rangle \\
\left| {{y_{output}}} \right\rangle  = \mathop {\hat F} {\Sigma _{new}}\mathop {{\hat w_{new}}} {\kern 1pt} \left| {{x_4}} \right\rangle  = \left| {011} \right\rangle  = {\kern 1pt} \left| {{y_4}} \right\rangle \\
\left| {{y_{output}}} \right\rangle  = \mathop {\hat F} {\Sigma _{new}}\mathop {{\hat w_{new}}} {\kern 1pt} \left| {{x_5}} \right\rangle  = \left| {100} \right\rangle  = \left| {{y_5}} \right\rangle \\
\left| {{y_{output}}} \right\rangle  = \mathop {\hat F} {\Sigma _{new}}\mathop {{\hat w_{new}}} {\kern 1pt} \left| {{x_6}} \right\rangle  = \left| {110} \right\rangle  = \left| {{y_6}} \right\rangle \\
\left| {{y_{output}}} \right\rangle  = \mathop {\hat F}{\Sigma _{new}}\mathop {{\hat w_{new}}} \left| {{x_7}} \right\rangle  = \left| {101} \right\rangle  = \left| {{y_7}} \right\rangle \\
\left| {{y_{output}}} \right\rangle  = \mathop {\hat F} {\Sigma _{new}}\mathop {{\hat w_{new}}}  \left| {{x_8}} \right\rangle  = \left| {111} \right\rangle  = \left| {{y_8}} \right\rangle \\
\left| {{y_{output}}} \right\rangle  = \mathop {\hat F}{\Sigma _{new}}\mathop {{\hat w_{new}}} \left| {{x_9}} \right\rangle  = \frac{{2\left| {010} \right\rangle  - 7\left| {110} \right\rangle }}{{\sqrt {{2^2} + {7^2}} }} = \left| {{y_9}} \right\rangle
\end{array}
\end{equation}
\textbf{Example 12} (less-complete) Suppose
\begin{equation}
\nonumber
\begin{array}{l}
\left| {{x_1}} \right\rangle  = \left| {010} \right\rangle ,{\kern 1pt} {\kern 1pt} {\kern 1pt} {\kern 1pt} {\kern 1pt} {\kern 1pt} {\kern 1pt} \left| {{y_1}} \right\rangle  = \left| {010} \right\rangle \\
\left| {{x_2}} \right\rangle  = \left| {101} \right\rangle ,{\kern 1pt} {\kern 1pt} {\kern 1pt} {\kern 1pt} {\kern 1pt} {\kern 1pt} {\kern 1pt} \left| {{y_2}} \right\rangle  = \left| {110} \right\rangle \\
\left| {{x_3}} \right\rangle  = \frac{{a\left| {000} \right\rangle  + b\left| {110} \right\rangle }}{{\sqrt {{a^2} + {b^2}} }},{\kern 1pt} {\kern 1pt} {\kern 1pt} {\kern 1pt} {\kern 1pt} {\kern 1pt} {\kern 1pt} \left| {{y_3}} \right\rangle  = \frac{{a\left| {000} \right\rangle  + b\left| {101} \right\rangle }}{{\sqrt {{a^2} + {b^2}} }}(a = 3,b = 2)
\end{array}
\end{equation}
\textit{Perceptron training}: The quantum perceptron is
\begin{equation}
\nonumber
\begin{array}{*{20}{l}}
{\left| {{y_{output}}} \right\rangle  = \mathop {{\rm{ }}\hat F} {\Sigma _{new}}\mathop {{\hat w_{new}}} \left| {{x_j}} \right\rangle }\\
\begin{array}{l}
= \frac{1}{{\sqrt {13} }}\left( {\begin{smallmatrix}
0&{ - 3}&0&2&0&0&0&0\\
0&0&0&0&0&{\sqrt {13} }&0&0\\
{\sqrt {13} }&0&0&0&0&0&0&0\\
0&0&0&0&0&0&{\sqrt {13} }&0\\
0&0&0&0&0&0&0&{\sqrt {13} }\\
0&{ - 2}&0&{ - 3}&0&0&0&0\\
0&0&{ - \sqrt {13} }&0&0&0&0&0\\
0&0&0&0&{\sqrt {13} }&0&0&0
\end{smallmatrix}} \right) \times \\
{\kern 8pt} {\kern 8pt} {\kern 1pt} {\kern 1pt}{\kern 1pt} {\kern 1pt}{\kern 1pt} {\kern 1pt} {\kern 1pt} \left( {\begin{smallmatrix}
1&0&0&0&0&0&0&0\\
0&1&0&0&0&0&0&0\\
0&0&1&0&0&0&0&0\\
0&0&0&1&0&0&0&0\\
0&0&0&0&1&0&0&0\\
0&0&0&0&0&1&0&0\\
0&0&0&0&0&0&1&0\\
0&0&0&0&0&0&0&1
\end{smallmatrix}} \right) \times \\
{\kern 1pt} {\kern 1pt}{\kern 1pt} {\kern 1pt}{\kern 1pt} {\kern 1pt}{\kern 1pt} {\kern 1pt} \frac{1}{{\sqrt {13} }}\left( {\begin{smallmatrix}
0&0&{\sqrt {13} }&0&0&0&0&0\\
{ - 3}&0&0&0&0&0&{ - 2}&0\\
0&0&0&0&0&{ - \sqrt {13} }&0&0\\
0&{\sqrt {13} }&0&0&0&0&0&0\\
0&0&0&0&0&0&0&{\sqrt {13} }\\
{ - 2}&0&0&0&0&0&3&0\\
0&0&0&{\sqrt {13} }&0&0&0&0\\
0&0&0&0&{\sqrt {13} }&0&0&0
\end{smallmatrix}} \right)\left| {{x_j}} \right\rangle
\end{array}
\end{array}
\end{equation}
\textit{Validation of correctness}:
\begin{equation}
\nonumber
\begin{array}{l}
\left| {{y_{output}}} \right\rangle  = \mathop {\hat F} {\Sigma _{new}}\mathop {{\hat w_{new}}} \left| {{x_1}} \right\rangle  = \left| {010} \right\rangle  = \left| {{y_1}} \right\rangle \\
\left| {{y_{output}}} \right\rangle  = \mathop {\hat F} {\Sigma _{new}}\mathop {{\hat w_{new}}} \left| {{x_2}} \right\rangle  = \left| {110} \right\rangle  = \left| {{y_2}} \right\rangle \\
\left| {{y_{output}}} \right\rangle  = \mathop {\hat F} {\Sigma _{new}}\mathop {{\hat w_{new}}} \left| {{x_3}} \right\rangle  = \frac{{3\left| {000} \right\rangle  + 2\left| {101} \right\rangle }}{{\sqrt {13} }} = \left| {{y_3}} \right\rangle
\end{array}
\end{equation}

\section{Performance evaluation}
\label{sec:5}
Now it is worth discussing the performance of the proposed algorithm. We select Altaisky's algorithm\cite{Altaisky01}, Seow et al's algorithm\cite{Seow15} as references, and evaluate our algorithm in the aspects of the training set, output result, and iteration times.

It is well known that training set used for a perceptron algorithm are not necessarily ideal, so the cases of the non-ideal training set, i.e., the less-complete and over-complete training sets, should be taken into account in practical scenarios. In Refs.\cite{Altaisky01} and \cite{Seow15}, they only consider the ideal case (i.e., complete training set), and it cannot be applied to those cases of less-complete and over-complete training sets. Table~\ref{tab:2} detailedly compares the availabilities of Altaisky's, Seow et al's and our algorithms. As shown in Table~\ref{tab:2}, our algorithm has higher availability than other two algorithms, i.e., it is not only suitable for a complete training set, but also for the less-complete and over-complete training sets.

\begin{table}[!t]
\centering
\footnotesize
\caption{Availability comparison among Altaisky's, Seow et al's and our algorithms under the complete, less-complete and over-complete training sets}
\label{tab:2}
\scalebox{1}[1]{
\tabcolsep 5pt 
\begin{tabular}{m{1.5cm}<{\centering}m{2.1cm}<{\centering}m{1.1cm}<{\centering}m{1.1cm}<{\centering}p{1.1cm}<{\centering}}
\toprule
   Quantum gates &Training set  &Altaisky's \cite{Altaisky01} & Seow et al's \cite{Seow15}& Ours    \\\hline
  \multirow{3}*{$H$} & complete set  & $\surd$ & $\surd$ & $\surd$\\
  ~ & less-complete set  & $\times$& $\times$ & $\surd$  \\
   ~ & over-complete set  & $\times$& $\times$ & $\surd$  \\\hline
\multirow{3}*{${S}$} & complete set & $\surd$& $\surd$ & $\surd$ \\
  ~ & less-complete set & $\times$& $\times$ & $\surd$ \\
  ~ & over-complete set & $\times$& $\times$ & $\surd$ \\\hline
\multirow{3}*{$T$} & complete set & $\surd$& $\surd$ & $\surd$ \\
  ~ & less-complete set & $\times$& $\times$ & $\surd$ \\
  ~ & over-complete set & $\times$& $\times$ & $\surd$ \\\hline
\multirow{3}*{\textit{CNOT}} & complete set & $\surd$&$\surd$ &$\surd$ \\
  ~ & less-complete set & $\times$& $\times$ & $\surd$ \\
  ~ & over-complete set & $\times$& $\times$ & $\surd$ \\\hline
\multirow{3}*{\textit{Toffoli}} & complete set & $\surd$& $\times$ & $\surd$\\
  ~ & less-complete set & $\times$& $\times$ & $\surd$ \\
  ~ & over-complete set & $\times$& $\times$ & $\surd$ \\\hline
\multirow{3}*{\textit{Fredkin}} & complete set & $\surd$&$\surd$ & $\surd$\\
  ~ & less-complete set & $\times$& $\times$ & $\surd$ \\
  ~ & over-complete set & $\times$& $\times$ & $\surd$ \\
\bottomrule
\end{tabular}
}
\end{table}

Assume that the training set is complete, we select six quantum gates $\{H, S, T, \textit{CNOT, Toffoli, Fredkin}\}$ to compare the accuracy (i.e., the difference between desirable results and output results) and applicability (i.e., applicable quantum gates) among three algorithms. Since Altaisky's algorithm converges to an approximate value after 30 iterations, so we take 30 as the iteration times, and the approximate value as the final output result. Different from Altaisky's algorithm, Seow et al's algorithm can get the accurate output results of five gates $\{H, S, T, \textit{CNOT, Fredkin}\}$ after only one iteration. In our algorithm, we also implement the \textit{Toffoli} gate within one iteration. As shown in Table~\ref{tab:3}, Seow et al's and our proposed algorithm have advantages over Altaisky's algorithm in terms of accuracy and iteration times, and our algorithm is more widely applicable than Seow's algorithm.

\begin{table}[!t]
\centering
\footnotesize
\caption{Comparison of iteration times and output results among Altaisky's, Seow et al's and our algorithms}
\label{tab:3}
\scalebox{1}[1]{
\begin{tabular}{m{2cm}<{\centering}m{2cm}<{\centering}m{2cm}<{\centering}m{2cm}<{\centering}m{2cm}<{\centering}}
\toprule
Quantum gates & Comparative content  &Altaisky's \cite{Altaisky01}& Seow et al's \cite{Seow15}& Ours    \\\hline
  \multirow{2}*{$H$} & Iteration times  & 30 & 1 & 1\\
   ~ & Output  & $\left( \begin{smallmatrix} 0.6771 \\ -0.7195 \end{smallmatrix} \right) $ & $\left( {\begin{smallmatrix}
{\frac{1}{{\sqrt 2 }}}\\
{\frac{1}{{\sqrt 2 }}}
\end{smallmatrix}} \right)$ & $\left( {\begin{smallmatrix}
{\frac{1}{{\sqrt 2 }}}\\
{\frac{1}{{\sqrt 2 }}}
\end{smallmatrix}} \right)$  \\\hline
\multirow{2}*{$S$} & Iteration times & 30 & 1 & 1 \\
  ~ & Output & $\left( {\begin{smallmatrix}0\\{0.9576i}
\end{smallmatrix}} \right)$ & $\left( {\begin{smallmatrix}
0\\{i}\end{smallmatrix}} \right)$ & $\left( {\begin{smallmatrix}
0\\{i}\end{smallmatrix}} \right)$ \\\hline
\multirow{2}*{$T$} & Iteration times & 30 & 1 & 1 \\
  ~ & Output & $\left( {\begin{smallmatrix}
0\\
{0.9576{e^{\frac{{i\pi }}{4}}}}
\end{smallmatrix}} \right)$ & $\left( {\begin{smallmatrix}
0\\
{{e^{\frac{{i\pi }}{4}}}}
\end{smallmatrix}} \right)$ & $\left( {\begin{smallmatrix}
0\\
{{e^{\frac{{i\pi }}{4}}}}
\end{smallmatrix}} \right)$ \\\hline
\multirow{2}*{\textit{CNOT}} & Iteration times & 30 & 1 & 1 \\
  ~ & Output & $\left( {\begin{smallmatrix}
0\\
0\\
{{\rm{0}}{\rm{.9576}}}\\
{{\rm{ - 0}}{\rm{.0424}}}
\end{smallmatrix}} \right)$ & $\left( {\begin{smallmatrix}
0\\
0\\
1\\
0
\end{smallmatrix}} \right)$& $\left( {\begin{smallmatrix}
0\\
0\\
1\\
0
\end{smallmatrix}} \right)$ \\\hline
\multirow{2}*{\textit{Toffoli}} & Iteration times & 30 & $-$ & 1\\
  ~ & Output & $\left( {\begin{smallmatrix}
0\\
0\\
0\\
0\\
0\\
0\\
{{\rm{0}}{\rm{.9576}}}\\
{{\rm{ - 0}}{\rm{.0424}}}
\end{smallmatrix}} \right)$ & $-$& $\left( {\begin{smallmatrix}
0\\
0\\
0\\
0\\
0\\
0\\
{\rm{1}}\\
{\rm{0}}
\end{smallmatrix}} \right)$ \\\hline
\multirow{2}*{\textit{Fredkin}} & Iteration times & 30 & 1 & 1\\
  ~ & Output & $\left( {\begin{smallmatrix}
0\\
0\\
0\\
0\\
0\\
0\\
0\\
{0.9576}
\end{smallmatrix}} \right)$& $\left( {\begin{smallmatrix}
0\\
0\\
0\\
0\\
0\\
0\\
{\rm{0}}\\
{\rm{1}}
\end{smallmatrix}} \right)$& $\left( {\begin{smallmatrix}
0\\
0\\
0\\
0\\
0\\
0\\
{\rm{0}}\\
{\rm{1}}
\end{smallmatrix}} \right)$ \\
\bottomrule
\end{tabular}
}
\end{table}

In summary, considering the aspects of availability, accuracy, and applicability, our algorithm has advantages over Altaisky's and Seow et al's algorithms.

\section{Discussion and conclusion}
\label{sec:6}
In this paper, we present an efficient one-iteration perceptron algorithm based on unitary weights. Different from the previous quantum perceptron algorithms, our proposed algorithm has good availability, high accuracy, and wide applicability. To be specific, our perception algorithm is not only suitable for the less-complete and over-complete training set, but also can accurately implement universal quantum gates within one iteration. In order to further validate its applicability, Appendix A gives an example to show that our algorithm can be applied to an arbitrary quantum composite gate (which can be viewed as arbitrary quantum computation).

The proposed algorithm discussed in our paper is a learning algorithm for the single-layer perceptron. For multi-layer perceptrons, where a hidden layer exists, more sophisticated algorithms such as backpropagation must be used. So, one of our next research efforts is how to use quantum technology to solve some problem of multi-layer perceptron. On the other hand, we know that when multiple perceptrons are combined in an artificial neural network, each output neuron runs independently of all other neurons; therefore, each output can be considered in isolation. Therefore, another work we would like to carry out is to select some neural network models and validate whether our quantum perceptron can be effectively used to construct a large-scale quantum neural network.

\section*{Acknowledgment}
The authors would like to express heartfelt gratitude to the anonymous reviewers and editor for their comments that improved the quality of this paper. And the support of all the members of the quantum research group of NUIST is especially acknowledged, their professional discussions and advice have helped us a lot. This work is supported by National Natural Science Foundation of China (Grant Nos. 61502101, 61501247, 61672290 and 71461005), Natural Science Foundation of Jiangsu Province (Grant Nos. BK20171458), the Six Talent Peaks Project of Jiangsu Province (Grant No. 2015-XXRJ-013), and the Priority Academic Program Development of Jiangsu Higher Education Institutions(PAPD).

\appendix
\renewcommand{\appendixname}{Appendix~\Alph{section}}
\section{Training arbitrary quantum composite gate with our algorithm}
\begin{figure}
\centering
\includegraphics[height=2cm]{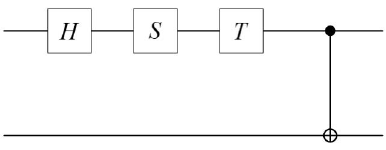}
\caption{Quantum circuit of the composite gate.}
\label{fig:5}
\end{figure}

In order to validate that the proposed quantum perceptron algorithm is suitable for the multiple gates composite calculation, the \textit{Hadamard} gate ($H$), \textit{phase} gate ($S$), $\pi /8$ gate ($T$), \textit{CNOT} gate are combined to construct the composite gate (shown in Fig.~\ref{fig:5}). Through the training of composite gate to further validate the applicability of our algorithm.

\noindent\textbf{Example 13} (over-complete) Suppose
\begin{equation}
\nonumber
\begin{array}{*{20}{l}}
{\kern 1pt} {\kern 1pt} {\kern 1pt} {\kern 1pt} {\left| {{x_1}} \right\rangle  = \left| {00} \right\rangle ,\left| {{y_1}} \right\rangle  = \frac{{\left| {00} \right\rangle  + i{e^{i\pi /4}}\left| {11} \right\rangle }}{{\sqrt 2 }}}\\
{\kern 1pt} {\kern 1pt} {\kern 1pt} {\kern 1pt} {\left| {{x_2}} \right\rangle  = \left| {01} \right\rangle ,\left| {{y_2}} \right\rangle  = \frac{{\left| {01} \right\rangle  + i{e^{i\pi /4}}\left| {10} \right\rangle }}{{\sqrt 2 }}}\\
{\kern 1pt} {\kern 1pt} {\kern 1pt} {\kern 1pt} {\left| {{x_3}} \right\rangle  = \left| {10} \right\rangle ,\left| {{y_3}} \right\rangle  = \frac{{\left| {00} \right\rangle  - i{e^{i\pi /4}}\left| {11} \right\rangle }}{{\sqrt 2 }}}\\
{\kern 1pt} {\kern 1pt} {\kern 1pt} {\kern 1pt} {\left| {{x_4}} \right\rangle  = \left| {11} \right\rangle ,\left| {{y_4}} \right\rangle  = \frac{{\left| {01} \right\rangle  - i{e^{i\pi /4}}\left| {10} \right\rangle }}{{\sqrt 2 }}}\\
\begin{array}{l}
\left| {{x_5}} \right\rangle  = \frac{{a\left| {00} \right\rangle  + b\left| {10} \right\rangle }}{{\sqrt {{a^2} + {b^2}} }},\\
\left| {{y_5}} \right\rangle  = \frac{{(a + b)\left| {00} \right\rangle  + (a - b)i{e^{i\pi /4}}\left| {11} \right\rangle }}{{\sqrt {2({a^2} + {b^2})} }}(a = 6,b = 8).
\end{array}
\end{array}
\end{equation}
\textit{Perceptron training}: The quantum perceptron is
\begin{equation}
\nonumber
\begin{array}{*{20}{l}}
{\left| {{y_{output}}} \right\rangle  = \mathop {\hat F} {\Sigma _{new}}\mathop {{\hat w_{new}}} \left| {{x_j}} \right\rangle }\\
{\begin{array}{*{20}{l}}
{{\kern 1pt} {\kern 1pt} {\kern 1pt} {\kern 1pt}  = \frac{1}{{10\sqrt 2 }}\left( {\begin{array}{*{20}{l}}
{ - 14}&2&0&0\\
0&0&{10}&{10}\\
0&0&{ - 10i{e^{i\pi /4}}}&{10i{e^{i\pi /4}}}\\
{2i{e^{i\pi /4}}}&{14i{e^{i\pi /4}}}&0&0
\end{array}} \right) \times }\\
{{\kern 1pt} {\kern 1pt} {\kern 1pt}{\kern 1pt} {\kern 1pt} {\kern 1pt} {\kern 1pt} {\kern 1pt} \left( {\begin{array}{*{20}{l}}
1&0&0&0\\
0&1&0&0\\
0&0&1&0\\
0&0&0&1
\end{array}} \right)\left( {\begin{array}{*{20}{l}}
{ - 0.6}&0&{ - 0.8}&0\\
{0.8}&0&{ - 0.6}&0\\
0&0&0&1\\
0&1&0&0
\end{array}} \right)\left| {{x_j}} \right\rangle }
\end{array}}
\end{array}
\end{equation}
\textit{Validation of correctness}:
\begin{equation}
\nonumber
\begin{array}{l}
\left| {{y_{output}}} \right\rangle  = \mathop {\hat F} {\Sigma _{new}}\mathop {{\hat w_{new}}} \left| {{x_1}} \right\rangle  = \frac{{\left| {00} \right\rangle  + i{e^{i\pi /4}}\left| {11} \right\rangle }}{{\sqrt 2 }} = \left| {{y_1}} \right\rangle \\
\left| {{y_{output}}} \right\rangle  = \mathop {\hat F} {\Sigma _{new}}\mathop {{\hat w_{new}}} \left| {{x_2}} \right\rangle  = \frac{{\left| {01} \right\rangle  + i{e^{i\pi /4}}\left| {10} \right\rangle }}{{\sqrt 2 }} = \left| {{y_2}} \right\rangle \\
\left| {{y_{output}}} \right\rangle  = \mathop {\hat F} {\Sigma _{new}}\mathop {{\hat w_{new}}} \left| {{x_3}} \right\rangle  = \frac{{\left| {00} \right\rangle  - i{e^{i\pi /4}}\left| {11} \right\rangle }}{{\sqrt 2 }} = \left| {{y_3}} \right\rangle \\
\left| {{y_{output}}} \right\rangle  = \mathop {\hat F} {\Sigma _{new}}\mathop {{\hat w_{new}}} \left| {{x_4}} \right\rangle  = \frac{{\left| {01} \right\rangle  - i{e^{i\pi /4}}\left| {10} \right\rangle }}{{\sqrt 2 }} = \left| {{y_4}} \right\rangle \\
\left| {{y_{output}}} \right\rangle  = \mathop {\hat F} {\Sigma _{new}}\mathop {{\hat w_{new}}} \left| {{x_5}} \right\rangle  = \frac{{14\left| {00} \right\rangle  - 2i{e^{i\pi /4}}\left| {11} \right\rangle }}{{\sqrt {2 \times 100} }} = \left| {{y_5}} \right\rangle
\end{array}
\end{equation}
\noindent\textbf{Example 14} (less-complete) Suppose
\begin{equation}
\nonumber
\begin{array}{*{20}{l}}
{\kern 1pt} {\kern 1pt} {\kern 1pt} {\kern 1pt} {\left| {{x_1}} \right\rangle  = \left| {00} \right\rangle ,\left| {{y_1}} \right\rangle  = \frac{{\left| {00} \right\rangle  + i{e^{i\pi /4}}\left| {11} \right\rangle }}{{\sqrt 2 }}}\\
\begin{array}{l}
\left| {{x_2}} \right\rangle  = \frac{{a\left| {01} \right\rangle  + b\left| {11} \right\rangle }}{{\sqrt {{a^2} + {b^2}} }},\\
\left| {{y_2}} \right\rangle  = \frac{{(a + b)\left| {01} \right\rangle  + (a - b)i{e^{i\pi /4}}\left| {10} \right\rangle }}{{\sqrt {2({a^2} + {b^2})} }}(a = 2,b = 1)
\end{array}
\end{array}
\end{equation}
\textit{Perceptron training}: The quantum perceptron is
\begin{equation}
\nonumber
\begin{array}{*{20}{l}}
{\left| {{y_{output}}} \right\rangle  = \mathop {{\rm{ }}\hat F} {\Sigma _{new}}\mathop {{\hat w_{new}}} \left| {{x_j}} \right\rangle }\\
\begin{array}{l}
 {\kern 1pt} {\kern 1pt} {\kern 1pt} {\kern 1pt} {\kern 1pt} {\kern 1pt} {\kern 1pt}  = \frac{1}{{\sqrt {10} }}\left( {\begin{array}{*{20}{l}}
{ - \sqrt 5 }&0&0&{\sqrt 5 {e^{i\pi /4}}}\\
0&{ - 3}&{{e^{i\pi /4}}}&0\\
0&{ - i{e^{i\pi /4}}}&3&0\\
{ - \sqrt 5 i{e^{i\pi /4}}}&0&0&{\sqrt 5 }
\end{array}} \right) \times \\
 {\kern 1pt}{\kern 1pt} {\kern 1pt}{\kern 1pt} {\kern 1pt} {\kern 1pt} {\kern 1pt} {\kern 1pt} {\kern 1pt} {\kern 1pt} {\kern 1pt} {\kern 1pt} {\kern 1pt} \left( {\begin{array}{*{20}{l}}
1&0&0&0\\
0&1&0&0\\
0&0&1&0\\
0&0&0&1
\end{array}} \right)\sqrt 5 \left( {\begin{array}{*{20}{l}}
{ - \sqrt 5 }&0&0&0\\
0&{ - 2}&0&{ - 1}\\
0&{ - 1}&0&2\\
0&0&{\sqrt 5 }&0
\end{array}} \right)\left| {{x_j}} \right\rangle
\end{array}
\end{array}
\end{equation}
\textit{Validation of correctness}:
\begin{equation}
\nonumber
\begin{array}{l}
\left| {{y_{output}}} \right\rangle  = \mathop {\hat F} {\Sigma _{new}}\mathop {{\hat w_{new}}} \left| {{x_1}} \right\rangle  = \frac{{\left| {00} \right\rangle  + i{e^{i\pi /4}}\left| {11} \right\rangle }}{{\sqrt 2 }} = \left| {{y_1}} \right\rangle \\
\left| {{y_{output}}} \right\rangle  = \mathop {\hat F} {\Sigma _{new}}\mathop {{\hat w_{new}}} \left| {{x_2}} \right\rangle  = \frac{{3\left| {01} \right\rangle {\rm{ + }}i{e^{i\pi /4}}\left| {10} \right\rangle }}{{\sqrt {2 \times 5} }} = \left| {{y_2}} \right\rangle
\end{array}
\end{equation}


\begin{thebibliography}{00}

\bibitem{Benioff80} P. Benioff, ``he computer as a physical system: A microscopic quantum mechanical Hamiltonian model of computers as represented by Turing machines,'' Journal of Statistical Physics, vol. 22, no. 5, pp. 563-591, May, 1980.

\bibitem{Feynman82} R. P. Feynman, ``Simulating physics with computers,'' International journal of theoretical physics, vol. 21, no. 6-7, pp. 467-488, 1982.

\bibitem{Deutsch85} D. Deutsch, ``Quantum theory, the Church¨CTuring principle and the universal quantum computer,'' Proc. R. Soc. Lond. A, vol. 400, no. 1818, pp. 97-117, 1985.

\bibitem{Shor99} P. W. Shor, ``Polynomial-Time Algorithms for Prime Factorization and Discrete Logarithms on a Quantum Computer,'' SIAM Review, vol. 41, no. 2, pp. 303-332, 1999.

\bibitem{Grover97} L. K. Grover, ``Quantum Mechanics Helps in Searching for a Needle in a Haystack,'' Physical Review Letters, vol. 79, no. 2, pp. 325-328, 1997.

\bibitem{Huang17} W. Huang, Q. Su, B. Liu, Y.-H. He, F. Fan, and B.-J. Xu, ``Efficient multiparty quantum key agreement with collective detection,'' Scientific Reports, vol. 7, no. 1, pp. 15264, 2017.

\bibitem{Liu18} W. J. Liu, Y. Xu, C. N. Yang, P. P. Gao, and W. B. Yu, ``An Efficient and Secure Arbitrary N-Party Quantum Key Agreement Protocol Using Bell States,'' International Journal of Theoretical Physics, vol. 57, no. 1, pp. 195-207, 2018.

\bibitem{Liu09} W. J. Liu, H. W. Chen, T. H. Ma, Z. Q. Li, Z. H. Liu, and W. B. Hu, ``An efficient deterministic secure quantum communication scheme based on cluster states and identity authentication,'' Chinese Physics B, vol. 18, no. 10, pp. 4105, 2009.

\bibitem{Zhong18} J. Zhong, Z. Liu, and J. Xu, ``Analysis and Improvement of an Efficient Controlled Quantum Secure Direct Communication and Authentication Protocol,'' Computers, Materials \& Continua, vol. 57, no. 3, pp. 621-633, 2018.

\bibitem{Tan18} X. Tan, X. Li, and P. Yang, ``Perfect Quantum Teleportation via Bell States,'' Computers, Materials \& Continua, vol. 57, no. 3, pp. 495-503, 2018.

\bibitem{Liu15} W. J. Liu, Z. F. Chen, C. Liu, and Y. Zheng, ``Improved Deterministic N-To-One Joint Remote Preparation of an Arbitrary Qubit via EPR Pairs,'' International Journal of Theoretical Physics, vol. 54, no. 2, pp. 472-483, 2015.

\bibitem{Qu17} Z. Qu, S. Wu, M. Wang, L. Sun, and X. Wang, ``Effect of quantum noise on deterministic remote state preparation of an arbitrary two-particle state via various quantum entangled channels,'' Quantum Information Processing, vol. 16, no. 12, pp. 306, 2017.

\bibitem{Wu18} F. Wu, X. Zhang, W. Yao, Z. Zheng, L. Xiang, and W. Li, ``An Advanced Quantum-Resistant Signature Scheme for Cloud Based on Eisenstein Ring,'' Computers, Materials \& Continua, vol. 56, no. 1, pp. 19-34, 2018.

\bibitem{Qu18} Z. Qu, T. Zhu, J. Wang, and X. Wang, ``A Novel Quantum Stegonagraphy Based on Brown States,'' Computers, Materials \& Continua, vol. 56, no. 1, pp. 47-59, 2018.

\bibitem{Qu1802} Z. Qu, Z. Cheng, W. Liu, and X. Wang, ``A novel quantum image steganography algorithm based on exploiting modification direction,'' Multimedia Tools and Applications, Online, Doi: 10.1007/s11042-018-6476-5, 2018.

\bibitem{Liu17} W.-J. Liu, Z.-Y. Chen, S. Ji, H.-B. Wang, and J. Zhang, ``Multi-party Semi-quantum Key Agreement with Delegating Quantum Computation,'' International Journal of Theoretical Physics, vol. 56, no. 10, pp. 3164-3174, 2017.

\bibitem{Liu1802} W. Liu, Z. Chen, J. Liu, Z. Su, and L. Chi, ``Full-Blind Delegating Private Quantum Computation,'' Computers, Materials \& Continua, vol. 56, no. 2, pp. 211-223, 2018.

\bibitem{Lloyd13} S. Lloyd, M. Mohseni, and P. Rebentrost, ``Quantum algorithms for supervised and unsupervised machine learning,'' arXiv preprint, arXiv:1307.0411, 2013.

\bibitem{Liu1803}W. J. Liu, P. P. Gao, W. B. Yu, Z. G. Qu, and C. N. Yang, ``Quantum Relief algorithm,'' Quantum Information Processing, vol. 17, no. 10, pp. 280, 2018.

\bibitem{McCulloch43} W. S. McCulloch, and W. Pitts, ``A logical calculus of the ideas immanent in nervous activity,'' The bulletin of mathematical biophysics, vol. 5, no. 4, pp. 115-133, 1943.

\bibitem{Rosenblatt58} F. Rosenblatt, ``The perceptron: a probabilistic model for information storage and organization in the brain,'' Psychological review, vol. 65, no. 6, pp. 386-408, 1958.

\bibitem{Kak95} S. Kak, ``On quantum neural computing,'' Information Sciences, vol. 83, no. 3, pp. 143-160, 1995.

\bibitem{Lagaris97} I. E. Lagaris, A. Likas, and D. I. Fotiadis, ``Artificial neural network methods in quantum mechanics,'' Computer Physics Communications, vol. 104, no. 1, pp. 1-14, 1997.

\bibitem{Purushothaman97} G. Purushothaman, and N. B. Karayiannis, ``Quantum neural networks (QNNs): inherently fuzzy feedforward neural networks,'' IEEE Transactions on Neural Networks, vol. 8, no. 3, pp. 679-693, 1997.

\bibitem{Andrecut02} M. Andrecut, and M. K. Ali, ``A quantum neural network model,'' International Journal of Modern Physics C, vol. 13, no. 01, pp. 75-88, 2002.

\bibitem{Gupta01} S. Gupta, and R. K. P. Zia, ¡°Quantum Neural Networks,¡± Journal of Computer and System Sciences, vol. 63, no. 3, pp. 355-383, 2001.

\bibitem{Nielsen00} M. A. Nielsen, and I. L. Chuang, ``Quantum computation and quantum information,'' Cambridge University Press, Cambridge, 2000.

\bibitem{Schuld14} M. Schuld, I. Sinayskiy, and F. Petruccione, ``The quest for a Quantum Neural Network,'' Quantum Information Processing, vol. 13, no. 11, pp. 2567-2586, 2014.

\bibitem{Panella11} M. Panella, and G. Martinelli, ``Neural networks with quantum architecture and quantum learning,'' International Journal of Circuit Theory and Applications, vol. 39, no. 1, pp. 61-77, 2011.

\bibitem{Altaisky01} M. Altaisky, ``Quantum neural network,'' arXiv preprint, quant-ph/0107012, 2001.

\bibitem{Behrman99} E. C. Behrman, J. E. Steck, and S. R. Skinner, ``A spatial quantum neural computer,'' vol. 2, pp. 874-877, 1999.

\bibitem{Zhou07} R. Zhou, and Q. Ding, ``Quantum M-P Neural Network,'' International Journal of Theoretical Physics, vol. 46, no. 12, pp. 3209-3215, 2007.

\bibitem{Sagheer13} A. Sagheer, and M. Zidan, ``Autonomous quantum perceptron neural network,'' arXiv preprint, arXiv:1312.4149, 2013.

\bibitem{Siomau14} M. Siomau, ``A quantum model for autonomous learning automata,'' Quantum Information Processing, vol. 13, no. 5, pp. 1211-1221, 2014.

\bibitem{Seow15} K. L. Seow, E. Behrman, and J. Steck, ``Efficient learning algorithm for quantum perceptron unitary weights,'' arXiv preprint, arXiv:1512.00522, 2015.

\bibitem{Silva16} A. J. da Silva, T. B. Ludermir, and W. R. de Oliveira, ``Quantum perceptron over a field and neural network architecture selection in a quantum computer,'' Neural Networks, vol. 76, pp. 55-64, 2016.


\end{thebibliography}
\end{document}